\documentclass[a4paper,fleqn]{cas-dc}

\usepackage[utf8]{inputenc}
\usepackage[numbers, sort&compress]{natbib}
\let\printorcid\relax %Remove ORCID footnote
\usepackage{balance}
\usepackage{lineno,hyperref}
\usepackage{comment}
\usepackage{amsmath}
\usepackage{upgreek}
\usepackage{tikz}
\usepackage{gensymb}
\usepackage{soul}
\usepackage{multirow}
\usepackage[short]{optidef}
\usepackage{graphicx}
\usepackage{subfigure}
\usepackage{color}
\usepackage{graphics}
\usepackage{float}
\usepackage{tablefootnote}
\usepackage{tabularx}
\usepackage{algorithm,algorithmic}
\usepackage{subfigure}
\usepackage{framed} % Framing content
\usepackage{multicol} % Multiple columns environment

\usepackage{nomencl} % Nomenclature package
\makenomenclature
\renewcommand*\nompreamble{\begin{multicols}{2}}
\renewcommand*\nompostamble{\end{multicols}}

\usepackage{ifthen}

\renewcommand{\nomgroup}[1]{%
	\ifthenelse{\equal{#1}{C}}{\item[\textit{Constants}]}
	{%
		\ifthenelse{\equal{#1}{V}}{\item[\textit{Variables}]}
		{%
			\ifthenelse{\equal{#1}{I}}{\item[\textit{Indices and sets}]}
			{%
				\ifthenelse{\equal{#1}{A}}{\item[\textit{Abbreviations}]}{}}}}
}

\def\tsc#1{\csdef{#1}{\textsc{\lowercase{#1}}\xspace}}
\tsc{WGM}
\tsc{QE}
\graphicspath{{./figures/}}
\begin{document}
\let\WriteBookmarks\relax
\def\floatpagepagefraction{1}
\def\textpagefraction{.001}

\shorttitle{Energy Management System for Resilience-Oriented Operation  of Ship Power Systems}    

\shortauthors{T. Nguyen et al. }  

\title[mode = title]{Energy Management System for Resilience-Oriented Operation  of Ship Power Systems}


%\tnotetext[1]{{\color{red} This work is supported by .....}}

% Author 1 information
\author[1]{Thai-Thanh Nguyen}
%\cormark[1] 
\ead{tnguyen@clarkson.edu} 
%\credit{Conceptualization of this study, Methodology, Software}

% Author 2 information
\author[1]{Bang Le-Huy Nguyen}
%\cormark[1] 
\ead{nguyenbl@clarkson.edu}

% Author i information
\author[1]{Tuyen Vu}
\cormark[1] 
\ead{tvu@clarkson.edu} 
%\address[1]{Clarkson University, Potsdam, NY, USA}
%\credit{Conceptualization of this study, Methodology, Software}
\cortext[cor1]{Corresponding author} 

\address[1]{Clarkson University, Potsdam, NY, USA}
%\address[2]{National Renewable Energy Research Lab, Golden, CO, USA}

% Funding information
\nonumnote{{This material is based upon research supported by, or in part by, the U.S. Office of Naval Research under award number N00014-16-1-2956}}

% Main

\begin{abstract}
	This paper proposes an original energy management methodology for enhancing the resilience of ship power systems considering multiple types of energy storage systems, including battery energy storage systems (BESS) and supercapacitor energy storage systems (SCESS). The primary function of the proposed EMS is to maximize the load operability while taking ramp-rate characteristics of energy storage systems (ESS) and generators into account innovatively. Balancing BESS's state-of-charge (SoC) and prioritizing the SoC level of SCESS are two additional objectives of the proposed EMS to manage energy storage systems. The receding horizon optimization (RHO) technique is proposed to reduce the computational burden, making the proposed method feasible for real-time applications. An all-electric MVDC ship power system is used to evaluate the performance of the proposed methodology. Simulation studies and results demonstrate the effectiveness of the proposed method in managing the ESS to ensure the system's resilience under generation power shortage. In addition, the proposed RHO technique significantly reduces the computation burden seen in the FHO technique while maintaining an acceptable resilience performance.
	
\end{abstract}

\begin{highlights}
	\item A resilience-oriented energy management system of ship power systems is proposed.
	\item Multiple types of energy storage systems such as battery energy storage and supercapacitor energy storage systems are considered.
	\item Multi-objective optimization problem and gradient descent algorithm are proposed to balance the trade-off between multiple objectives. 
	\item A comparison study between the receding horizon optimization and fixed horizon optimization is presented. 
	\item Proposed receding horizon optimization for ship power systems enables real-time implementations. 
\end{highlights}


\maketitle

% Nomenclature Section --------------------

\begin{table*}[!t]
	\begin{framed}
		% Abbreviations
		\nomenclature[A]{AES}{All-electric ship}
		\nomenclature[A]{ESS}{Energy storage system}
		\nomenclature[A]{EMS}{Energy management system}
		\nomenclature[A]{SPS}{Ship power system}
		\nomenclature[A]{BESS}{Battery energy storage system}
		\nomenclature[A]{SCESS}{Super-capacitor energy storage system}
		\nomenclature[A]{RHO}{Receding horizon optimization}
		\nomenclature[A]{FHO}{Fixed horizon optimization}
		\nomenclature[A]{SoC}{State-of-charge}
		\nomenclature[A]{HRRL}{High ramp-rate load}
		\nomenclature[A]{PGM}{Power generation module}
		\nomenclature[A]{PCM-1A}{Power conversion module}
		\nomenclature[A]{IPNC}{Integrated power 			node center}
		\nomenclature[A]{PMM}{Propulsion motor module}
		\nomenclature[A]{ESM}{Energy storage module}
		
		% Constant
		\nomenclature[C]{$n_L$}{Number of loads}
		\nomenclature[C]{$n_E$}{Number of ESS}
		\nomenclature[C]{$n_G$}{Number of generators}
		\nomenclature[C]{$N_p$}{Otimization horizon length}
		\nomenclature[C]{$T$}{Mission time}
		\nomenclature[C]{$\hat{w}_i$}{Load weight}
		\nomenclature[C]{$\alpha_e$}{Weight of SoC}

		% Variables
		\nomenclature[V]{$o_i^t$}{Operation status of load}
		\nomenclature[V]{$P_e^{E,t}$}{ESS power}
		\nomenclature[V]{$P_g^{G,t}$}{Generator power}
		\nomenclature[V]{$u_P^t$}{Auxiliary variable of absolute ESS power}
		\nomenclature[V]{$u_{SoC}^t$}{Auxiliary variable of SoC difference}
		\nomenclature[V]{$r_e^{E,t}$}{Power ramp-rate of ESS}
		\nomenclature[V]{$r_g^{G,t}$}{Power ramp-rate of generator}
		% Indices and sets
		\nomenclature[I]{$t$}{Time index}
		\nomenclature[I]{$i$}{Load index}
		\nomenclature[I]{$e$}{ESS index}
		\nomenclature[I]{$g$}{Generator index}
		\printnomenclature
	\end{framed}
\end{table*}

\section{Introduction}

%Ship power system overview. 

Modern shipboard power systems (SPS) are shifting toward all-electric ships (AES), which integrate advanced systems such as electric propulsion, power conversion, energy storage systems (ESS), and intelligent management systems. Electrifying SPS enables the uses of ESS for improving fuel efficiency \cite{LAN201526, kanellos2013optimal, 7793272} and serving high power ramp-rate loads \cite{van2017predictive}. Due to the vulnerability of SPS to system failures such as tripping generations, ESS devices can be utilized as backup sources to compensate for power shortages caused by tripping generation units \cite{GEERTSMA201730, wen2016allocation}. 

% https://energyeducation.ca/encyclopedia/Energy_density_vs_power_density
ESS integrated into ships can be categorized into two types based on their characteristics. The first has high energy density but low power density, while the second has high power density but low energy density \cite{PAN2021111048}. Battery energy storage systems (BESS) belong to the first type, which can operate for a relatively long period of time. The second type includes super-capacitor energy storage systems (SCESS), which can generate high power for a short period of time. The uses of both types will offer benefits of high power density and high energy density. BESS can be used to serve base loads and moderate ramp-rate loads whereas SCESS serve critical high ramp-rate loads \cite{8638571, 9224178}. The uses of hybrid ESS, including BESS and SCESS in AES, have been presented in \cite{Khan7829349, faddel2019coordination, peng2019optimization}, in which fuzzy controls were used to manage hybrid ESS system for SPS with pulsed loads. Either low-pass filter \cite{Khan7829349} or high-pass filter \cite{faddel2019coordination} could be used to separate the low and high-frequency components, then the fuzzy controls are used to produce the reference signals for the BESS and SCESS accordingly. Model predictive controls of hybrid ESS for mitigating fluctuation in loads and increasing fuel efficiency have been presented in \cite{HOU201862, HASELTALAB2019113308}. Optimal power flow for ship power systems with hybrid ESS was presented in \cite{9314261}. Although these strategies deal with energy management of hybrid ESS, they might fail to manage SPS in conditions of generation shortages, such as tripping generators. In such conditions, load shedding control is required to maintain power balance. Effectively managing hybrid ESS will minimize load shedding amount and enhance system resilience over a full mission scenario. A resilience-oriented energy management system (EMS) considering different characteristics of ESS plays an important role in future SPS.

The objective of the resilience-oriented operation of ship power systems is to minimize the amount of load shedding due to the failures of generations or operation of high power ramp-rate load (HRRL). The importance of loads can be categorized into vital or non-vital loads, which are defined by the weight values. When the available power generation is less than load demands, the non-vital loads (those with the low weight values) should be shed first. Existing studies on the resilience enhancement of SPS can be divided into two categories: centralized and decentralized approaches. In the centralized approach, the optimization problem is solved by a single control center, whereas in the decentralized approach, it is solved by multiple control units. 

Centralized EMS for enhancing the resilience of ship power systems based on probabilistic methods have been presented in \cite{9131191, momoh2002optimal,9207909, 7055266}. Dynamic prioritization of the loads was considered for load shedding problem in \cite{ding2009dynamic, 4906545}. However, ESS were excluded from these studies. Optimal power management in \cite{xu2018optimal} performs load shedding and reconfiguration to minimize the impact of fault on ship power systems. A two-phase optimization problem to improve resilience was proposed in \cite{li2018resilience}, in which the first phase of the optimization problem maximizes load survivability, whereas the second phase maximizes the functionality of supplying loads. A graph-theoretic method considering line capacities and load priorities was used in \cite{lai2018graph} to enhance system resilience against physical attacks on power lines. Although ESS are also not involved in \cite{xu2018optimal, li2018resilience, lai2018graph}, the optimization problems in these studies are still complex and challenging to solve. Thus, constraint relaxation was utilized to formulate a new low-complexity problem that ensures feasible near-optimal solutions. With the inclusion of several types of ESS, the complexity of existing centralized optimization problems will increase significantly, which poses a considerable impact on the computational burden. 

Distributed strategies in \cite{lai2018distributed, edrington2020distributed, MOHAMED2020118041, 8847883, 6876221, 6197250} overcomes the computational limitation of the centralized approaches as the global optimization problem is solved by multiple controllers. In \cite{lai2018distributed, edrington2020distributed, MOHAMED2020118041, 8847883}, the ship power system was divided into multiple zones and the distributed EMS based on the alternating direction method of multipliers algorithm were used to solve the optimization in the distributed manner. In \cite{6876221, 6197250}, multi-agent approaches were presented to manage load shedding in ship power systems. However, limitations of these distributed EMS are the complexity of communication systems and the cyber-security issues due to their dependency on communication networks. The decentralized method in \cite{faddel2019decentralized} manages hybrid ESS based on local voltage profiles, avoiding the use of complex communication networks. However, decentralized EMS in \cite{faddel2019decentralized} neglects load shedding problem might put the ship power system at risk due to the potential insufficient generation capacity.  

Existing energy management approaches have their own advantages and disadvantages. The centralized methods have advantages of simplicity and high accuracy, but they have the drawback of being computationally intensive, especially when dealing with large numbers of variables \cite{faddel2019decentralized}. The decentralized methods overcome the computational limitation of the centralized method as the problem is solved in a distributed manner, but they face cyber-security issues due to complex communication networks \cite{9026756, 9205672, 6870484}. Existing studies on the resilience enhancement of ship power systems are either computationally intensive or facing cyber-security issues. To address the problem, this paper proposes a receding horizon optimization (RHO) for the resilience-oriented operation of the ship power systems. The proposed methodology has a computational benefit over existing centralized methods since the RHO solves the optimization problem with a sequence of short trajectories rather than a single large trajectory. The proposed method is comparable to the distributed methods in terms of computational effort, but it is simpler and easier to implement due to the absence of complex communication networks, thus reducing the risk of cyber attacks. In addition, existing studies have not considered the energy management of multiple ESS, such as balancing SoC among ESS or prioritizing the use of ESS, in addition to the main objective of resilience enhancement. Since the SoC estimation is not perfect, the long time operation would cause large SoC variances among ESS. Prioritizing specific type of ESS for a particular role is an important factor as different types of ESS have different characteristics. SoC balancing and ESS prioritizing are two secondary objectives being considered in the proposed EMS. The proposed methodology is evaluated in the four-zone notional MVDC system. A comparison study on the proposed RHO and FHO methods is also carried out. \\

\textbf{The main contributions of this paper are as follows:}
\begin{itemize}
	\item A resilience-oriented EMS is proposed for ship power systems that include multiple types of ESS, such as BESS and SCESS. 
	\item The proposed EMS not only enhances the system resilience but also manages ESS for long-term operation, such as balancing SoC among ESS and prioritizing the SoC level of SCESS.
	\item Receding horizon optimization is proposed to solve the optimization problem, which significantly reduces the computational burden, making it suitable for the real-time application.
\end{itemize} 

	The rest of this paper is organized as follows: Section~\ref{sec:problem} describes the resilience problem of ship power systems and formulating resilience enhancement into an optimization problem. The objective function and constraints are presented in this section. The proposed receding horizon optimization is presented in Section~\ref{sec:RHO}. Case studies of notional four-zone MVDC ship power systems are described in Section~\ref{sec: casestudies}. A comparison study on the RHO and FHO methods is also given in this section. Finally, the main findings of this paper are summarized in Section~\ref{sec:conclusion}.

\section{Problem Formulation}
\label{sec:problem}

The importance of loads is defined by a weight value $w_i$. For example, the vital load may have a weight of 1, and the non-vital load may have a weight of 0.1. The importance of loads can vary according to the mission operations, resulting in the variation of load's weight in different missions. Load operability ($O$) quantifies the degree of served loads, as given by (\ref{eq:Oi}), which is modified from \cite{4233795} to ensure that the importance of loads and corresponding rated power are involved in load shedding problems. The resilience of ship power systems is enhanced by optimally scheduling generators and ESS to minimize load shedding, thus maximizing load operability.

\begin{align}
	\label{eq:Oi}
	O  & = {{\int_{t_0}^{t_f}}{\sum_{i = 1}^{n_L}}{\hat{w}_i o_i^t} dt
		\over 
		{\int_{t_0}^{t_f}}{\sum_{i = 1}^{n_L}}{\hat{w}_i o_i^{*t}} dt
	}, \\
	\hat{w}_i & = w_i P_{Li}^{rated} & \forall i \in n_L,
\end{align}
where $n_L$ is the number of loads; $P_{Li}^{rated}$ is the rated power of load $i$; $\hat{w}_i$ is the normalized weight value of load $i$; $o_i^t$ is the operational status of load $i$ at time $t$; and $o_i^{*t}$ is the commanded operational status of load $i$; $t_0$ represents the start time of the event, while $t_f$ represents the end time of the event.

\begin{figure}[t]
	\centering
	\includegraphics[width=0.9\linewidth]{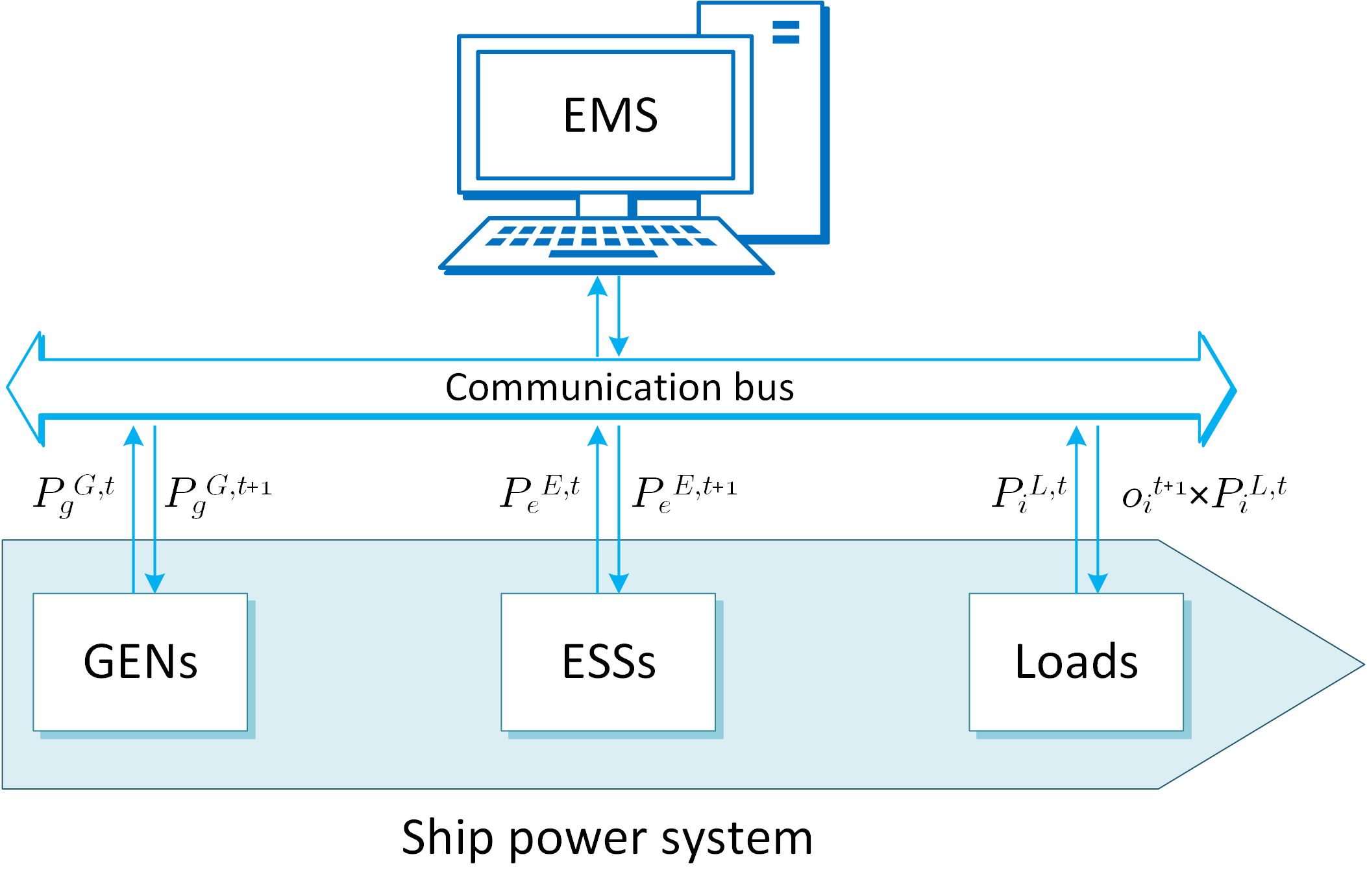}
	\caption{Energy management of ship power systems.}
	\label{fig:EMShip}
\end{figure}

Fig.~\ref{fig:EMShip} shows the EMS of the ship power system, including communication signals. The centralized EMS gathers all information of the generators (GEN), ESS, and loads then the optimization process is performed to find the operational status of loads and power of GEN and ESS. Those optimal values are sent back to the ship power system as the reference values. Local controllers of generators and ESS take actions to track the reference signals. Loads will be shed if the value of operational status is smaller than commanded operational status. 

\subsection{Objective Function}

The objective function for enhancing system resilience in the proposed EMS is given by (\ref{eq:objfunc}), which includes four terms, as shown in (\ref{eq:objfuncf1}) to (\ref{eq:objfuncf4}). The first term, $f_1(o_i^t)$, represents for load operability over mission time $T$. The second term, $f_2(P_e^{E,t})$, is used to minimize control actions of ESS. The third term, $f_3(SoC^t)$, minimizes the state-of-charge (SoC) difference between ESS. The final term, $f_4(SoC^{T})$, is used to maximize SoC at the end of the optimization window, which makes the proposed method distinguish from existing optimization algorithms that maintain the same SoC levels at initial and final intervals of the optimization window. The final term of objective function helps the receding horizon optimization maintaining SoC level for long-term operation. 

\begin{maxi}[1]
	{x}
	{	f(x) = f_1(o_i^t) 
			- \omega_1 f_2(P_e^{E,t}) 
	}
	{}{}
	\breakObjective{- \omega_2 f_3(SoC^t) + \omega_3 f_4(SoC^{T})},
	\label{eq:objfunc}
\end{maxi}
%where
\begin{align}
	x & = [o_i^t,P_e^{E,t}, P_g^{G,t}]^\top, \\
	\label{eq:objfuncf1}
	f_1(o_i^t)  
	&= {
		\sum_{t \in 1}^T	
		{\sum_{i \in 1}^{n_L}{\hat{w}_i o_i^t}}},\\
	f_2(P_e^{E,t})  
	\label{eq:objfuncf2}
	&= \sum_{t \in 1}^T{\sum_{e \in 1}^{n_E}{|P_e^{E,t}|}},\\
	f_3(SoC^t) 
	\label{eq:objfuncf3}
	&= \sum_{t \in 1}^T{\sum_{l,m}{|SoC_l^{E,t} - SoC_m^{E,t}|}},\\
	f_4(SoC^{T}) 
	\label{eq:objfuncf4}
	&= \sum_{e \in 1}^{n_E}{\alpha_e SoC_e^{E,T}},	
\end{align}
where $T$ is the mission time; $o_i^t$ is the operability of load $i$ at time $t$; $\hat{w}_i$ is the weight of load $i$, which indicates the importance of load such as critical or non critical loads; $P_e^{E,t}$ is the active power of ESS $e$ at time $t$; $SoC^t$ is the ESS's SoC at time $t$; and $SoC^T$ is the ESS's SoC at the end of optimization window; $\omega_1, \omega_2,$ and $\omega_3$ are the constant weights, which will be optimally selected by the gradient descent algorithm in the next section.

\subsection{Constraints}

Objective function (\ref{eq:objfunc}) is subjected to power supply-demand constraint given in (\ref{eq:dP}), as the optimal load operability has to be smaller than total generations including ESS power. 
\begin{align}
	\label{eq:dP}
	\sum_{i=1}^{n_L}{P_{i}^{L,t}} o_i^t \leq &
\sum_{e = 1}^{n_E}{P_e^{E,t}} + 
\sum_{g = 1}^{n_G}{P_g^{G,t}}
&{\forall t \in T},
\end{align}
where $n_L$ is the number of loads; $P_{i}^{L,t}$ is the command of load $i$ at time $t$; $n_E$ is the number of ESS; $n_G$ is the number of generations; and $P_g^G$ is the power of generator $g$.

Operational status $o_i^t$ is subjected to the limit constraints~(\ref{eq:oilim}), in which the commanded operational status is equal to one ($o_i^{*t} = 1$). If $o_i^t  = 0$, load $i$ must be shed 100\% at time~$t$; if $o_i^t  = 1$, load $i$ is served 100\% at time~$t$. In addition, ESS and generators are also subjected to the limit constraints in (\ref{eq:Pelim}) and (\ref{eq:Pglim}).
\begin{align}
	\label{eq:oilim}
	0& \leq o_i^t \leq 1&{\forall i \in n_L, \forall t \in T}, \\
	\label{eq:Pelim}
	P_e^{\min} & \leq P_e^{E,t} \leq P_e^{\max} &{\forall e \in n_E, \forall t \in T}, \\
	\label{eq:Pglim}
	P_g^{\min} & \leq P_g^{G,t} \leq P_g^{\max} & {\forall g \in n_G, \forall t \in T},
\end{align}
where $P_e^{\min}$ and $P_e^{\max}$ are the minimum and maximum capacity of ESS, respectively; $P_g^{\min}$ and $P_g^{\max}$ are the minimum and maximum capacity of generator, respectively.

Since some loads have discrete changes, the operational status of such loads is subjected to the discrete function, as given by (\ref{eq:oistep}). 
\begin{align}
	\label{eq:oistep}
	o_i^t \in 
	\begin{cases}
		\mathcal{Z} & \text{if load } i \text{ is the step load,}  \\
		[0, 1] & 	\text{otherwise},
	\end{cases}
\end{align}
where $\mathcal{Z} = \{0:\Delta o_i:1\}$; $\Delta o_i = 1/n$; $n \in \mathbb{Z}^+$ is the number of steps of the load command. 

Since both generators and ESS have limit on power ramp-rate, reference power set points to these devices is subjected to ramp-rate constraints (\ref{eq:rglim}) and (\ref{eq:relim}).
\begin{align}
	\label{eq:rglim}
	r_g^{\min} & \leq r_g^{G,t} \leq r_g^{\max} & {\forall g \in n_G, \forall t \in T}, \\
	\label{eq:relim}
	r_e^{\min} & \leq r_e^{E,t} \leq r_e^{\max} & {\forall e \in n_E, \forall t \in T}, \\
	r_g^{G,t} & = {P_g^{G,t} - P_g^{G,t-1} \over {\Delta t}} & {\forall g \in n_G, \forall t \in T}, \\
	r_e^{E,t} & = {P_e^{E,t} - P_e^{E,t-1} \over {\Delta t}} & {\forall e \in n_E, \forall t \in T},
\end{align}
where $\Delta t$ is the sampling time.

Finally, to prevent dip charge and discharge of ESS, the SoC level has to be in limit (\ref{eq:SoClim}). 
\begin{align}
	\label{eq:SoClim}
	SoC_e^{\min} & \leq SoC_e^{E,t} \leq SoC_e^{\max} & { \forall e \in n_E, \forall t \in T}, \\
	SoC_e^{E,t} & = SoC_e^{E,t-1} + \Delta t {P_e^{E,t} \over {P_e^{\max}}} & {\forall e \in n_E, \forall t \in T}.	
\end{align}

\subsection{MILP to Solve Optimization Problem}

The problem formulated from Section \ref{sec:problem} includes  discrete variable of operational status and non-linear objective terms ($f_2$ and $f_3$). To solve (\ref{eq:objfunc}) by the mixed-integer linear programming (MILP), the discrete variable is converted to integer variable and the non-linear objective terms are linearized. 

To convert discrete variable to integer variable, the load power ($P_{i}^{L,t}$) and load weight value ($\hat{w}_i$) of the discrete loads are scaled with the a factor of step size ($\Delta o_i$), as given in (\ref{eq:scaledwi}) and (\ref{eq:scaledPL}).
\begin{align}
	\label{eq:scaledwi}
	\hat{w}_i &= \hat{w}_i \Delta o_i \\
	\label{eq:scaledPL}
	\hat{P}_{i}^{L,t} & = {P_{i}^{L,t} \Delta o_i} 
\end{align}
where $\hat{w}_i$ and $\hat{P}_{i}^{L,t}$ are the modified load weight and power, respectively. Substituting (\ref{eq:scaledwi}) and (\ref{eq:scaledPL}) into (\ref{eq:objfuncf1}) and (\ref{eq:dP}), respectively, variable of operational status is converter to the integer variable subjected to the following constraint.
\begin{align}
	\label{eq:limoihat}
	0 \leq o_i^t \leq {1 \over \Delta o_i}
\end{align}

The optimal variables of operation status found by optimization process are converted back to discrete variables using (\ref{eq:scaledoi2}), which will be used as the load commands for the discrete loads.
\begin{align}
	\label{eq:scaledoi2}
	\hat{o}_i^t = { o_i^t \Delta o_i}
\end{align}

Non-linear objective terms are linearized by introducing auxiliary variables and constraints for such variables, as given by (\ref{eq:aux_uP}) and (\ref{eq:aux_uSoC}). 
\begin{align}
	\label{eq:aux_uP}
	u_P^t & = |P_e^{E,t}|, \\
	\label{eq:aux_uSoC}
	u_{SoC}^t & = |SoC_l^{E,t} - SoC_m^{E,t}|
\end{align}

Objective function (\ref{eq:objfunc}) can be represented by (\ref{eq:objfunc_linearize}) subjected to new constraints of operational status, power balance, and additional constraints of auxiliary variables. Thus, MILP can be used to solve the problem (\ref{eq:objfunc_linearize}).
\begin{maxi}[1]
	{x}
	{	f(x) = f_1(o_i^t) 
		- \omega_1 f_2(u_P^t) 
	}
	{}{}
	\breakObjective{- \omega_2 f_3(u_{SoC}^t) + \omega_3 f_4(SoC^{T})},
	\addConstraint{0& \leq o_i^t \leq {1 \over \Delta o_i}}{\forall i \in n_L, \forall t \in T} 
	\addConstraint{\sum_{i=1}^{n_L}{\hat{P}_{i}^{L,t}} o_i^t & \leq \sum_{e = 1}^{n_{E,t}}{P_e^E} + \sum_{g = 1}^{n_G}{P_g^{G,t}}} {\forall t \in T}
	\addConstraint{0 & \leq u_P^t \leq P_e^{\max}}{ \forall e \in n_E, \forall t \in T} 
	\addConstraint{u_P^t & \geq  P_e^{E,t}}{\forall e \in n_E, \forall t \in T} 
	\addConstraint{u_P^t & \geq  -P_e^{E,t}}{\forall e \in n_E, \forall t \in T} 
	\addConstraint{0 & \leq u_{SoC}^t \leq SoC_e^{\max}}{ \quad\forall e \in n_E, \forall t \in T} 
	\addConstraint{u_{SoC}^t & \geq SoC_l^{E,t} - SoC_m^{E,t} }{\forall l, m \in n_E, \forall t \in T} 
	\addConstraint{u_{SoC}^t & \geq -SoC_l^{E,t} + SoC_m^{E,t} }{\forall l, m \in n_E, \forall t \in T},
	\label{eq:objfunc_linearize}
\end{maxi}
where
\begin{align}
	x & = [o_i^t,P_e^{E,t}, P_g^{G,t}, u_P^t, u_{SoC}^t]^\top, \\
	\label{eq:objfuncf12}
	f_1(o_i^t)  
	&= {
		\sum_{t \in 1}^T	
		{\sum_{i \in 1}^{n_L}{\hat{w}_i o_i^t}}},\\
	f_2(u_P^t)  
	\label{eq:objfuncf2u}
	&= \sum_{t \in 1}^T{\sum_{e \in 1}^{n_E}{u_P^t}},\\
	f_3(u_{SoC}^t) 
	\label{eq:objfuncf3u}
	&= \sum_{t \in 1}^T{\sum_{l,m}{u_{SoC}^t}}.
\end{align}

\subsection{Gradient descent Algorithm to Optimize Objective Weights}

\begin{figure}[t]
	\centering
	\includegraphics[width=0.9\linewidth]{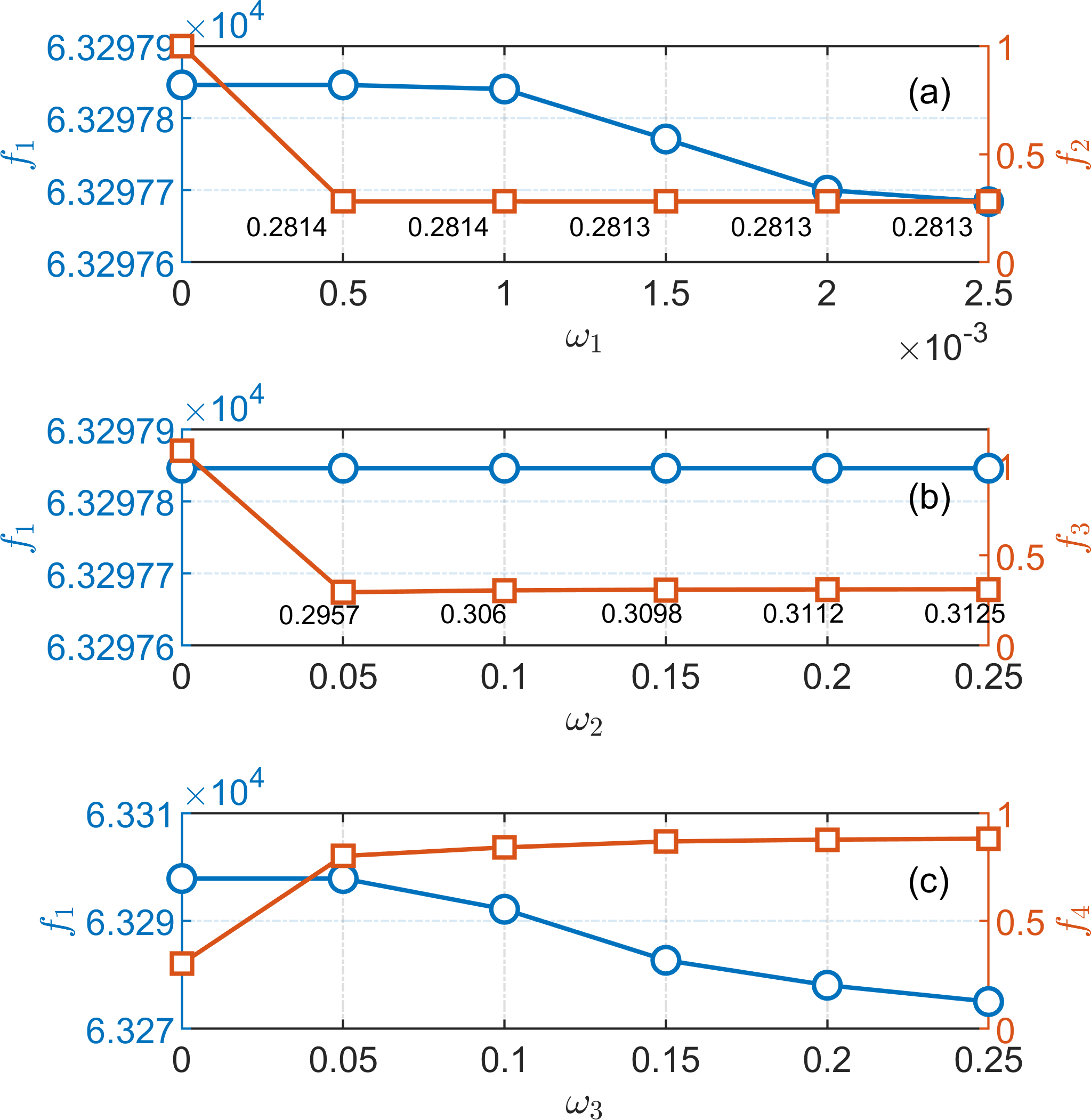}
	\caption{Effect of secondary objectives on load operability.}
	\label{fig:effectkgains}
\end{figure}

Constant weights, $\omega = [\omega_1, \omega_2, \omega_3]$, in objective function (\ref{eq:objfunc}) have an impact on load operability (\ref{eq:objfuncf1}), which is shown in Fig.~\ref{fig:effectkgains}. The increases of $\omega_1$ and $\omega_3$ result in the reduction of load operability. Because a high value of $\omega_1$ prevents ESS actions, which results in shedding more loads. Besides, a high value of $\omega_3$ makes ESS charge frequently to maximize SoC level, which also results in shedding more loads. On the other hand, constant weight $\omega_2$ has a slight impact on load operability as it tries to balance the SoC level among ESS.

From above observation, it can be seen that function $\bar{f}$ in (\ref{eq:GDfunc}) is convex. Gradient descent (GD) algorithm in \textbf{Algorithm}~1 is used to find constant weights $\omega_i$. GD stop iterations when absolute error of function $\bar{f}$ is smaller than a pre-defined value $\epsilon = 10^{-4}$. Optimal weights found by GD algorithm are $\omega = [0.0056; 0.0321; 0.0541]$.

\begin{align}
	\label{eq:GDfunc}
	\bar{f} = -\bar{f}_1 + \bar{f}_2 + \bar{f}_3 - \bar{f}_4
\end{align}
where $\bar{f_i}$ is the normalized function of $f_i$.

\begin{algorithm}[h]
	\begin{algorithmic}[1]
		\STATE $i \leftarrow 0$, initialize $\omega = [\omega_1; \omega_2; \omega_3]$
		\REPEAT 
		\STATE Calculate $\bar{f}$ based on $\omega_1$, $\omega_2$, and $\omega_3$: 
		$\bar{f}(i) \leftarrow \bar{f}|\omega$;
		
		\STATE $g(i) \leftarrow {\big[{\bar{f}(i) - \bar{f}(i-1)}\big] / \big[{\omega(i) - \omega(i-1)}\big]}$;
		
		%			\STATE $g(i) \leftarrow {{f(i) - f(i-1)} \over {\omega(i) - \omega(i-1)}}$;
		
		\STATE	$\omega(i+1) \leftarrow \omega(i) - \gamma g(i);$
		\STATE $i \leftarrow i+1$
		\UNTIL{$|\bar{f}(i+1) - \bar{f}(i)| < \epsilon$}
		\RETURN $\omega$;
	\end{algorithmic} 
	\label{alg:GDweights}
	\caption{Optimizing weights based on GD algorithm.}
\end{algorithm}

\begin{table}[t]
	\caption{System parameters}
	\label{TB:systemparameters}
	\begin{tabularx}{3.3in}{c|c|c}
		\toprule
		Symbol &  Parameter & Value\\ % Table header row
		\midrule
		$n_E$ 		&	Number of ESS		& 4 \\
		$P_E^{\max}$ 		&	ESS maximum power		& $10$~MW \\
		$P_E^{\min}$ 		&	ESS minimum power		& $-10$~MW \\
		$r_g^{\max,\min}$ 		&	Power ramp-rate limit of generator	& $\pm1$~MW/s \\
		$r_{BESS}^{\max,\min}$ 		&	Power ramp-rate limit of BESS	& $\pm5$~MW/s \\
		$r_{SCESS}^{\max,\min}$ 		&	Power ramp-rate limit of SCESS	& $\pm100$~MW/s \\
		$E_{BESS}$ 		&	Energy level of BESS	& $1000$~MJ \\
		$E_{SCESS}$ 		&	Energy level of SCESS	& $200$~MJ \\
		
		\bottomrule
	\end{tabularx}
\end{table}

\begin{figure}[t]
	\centering
	\includegraphics[width=0.9\linewidth]{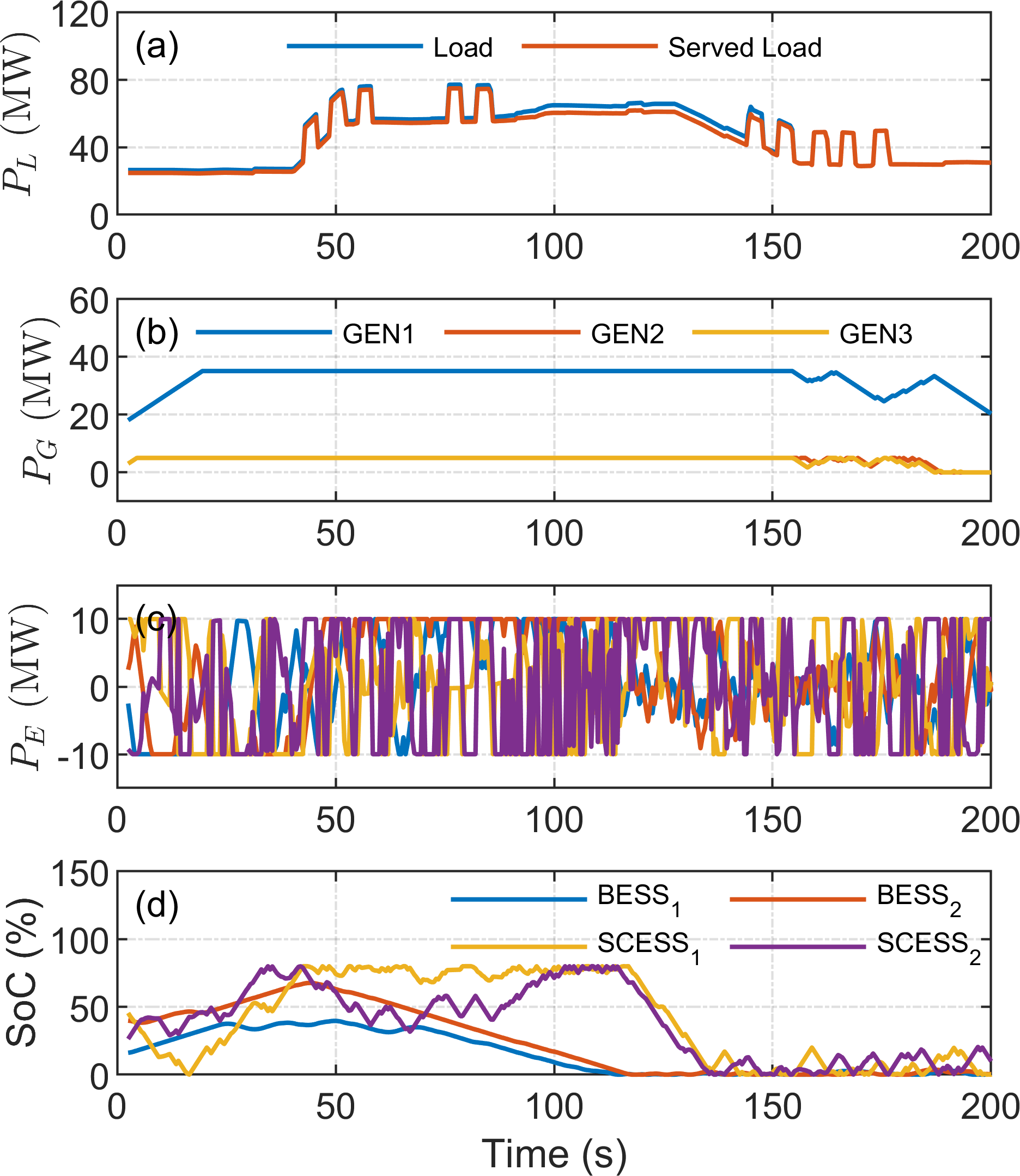}
	\caption{Optimization problem without secondary objectives ($f_1, f_2,$ and $f_3$).}
	\label{fig:LSwoGD}
\end{figure}

\begin{figure}[t]
	\centering
	\includegraphics[width=0.9\linewidth]{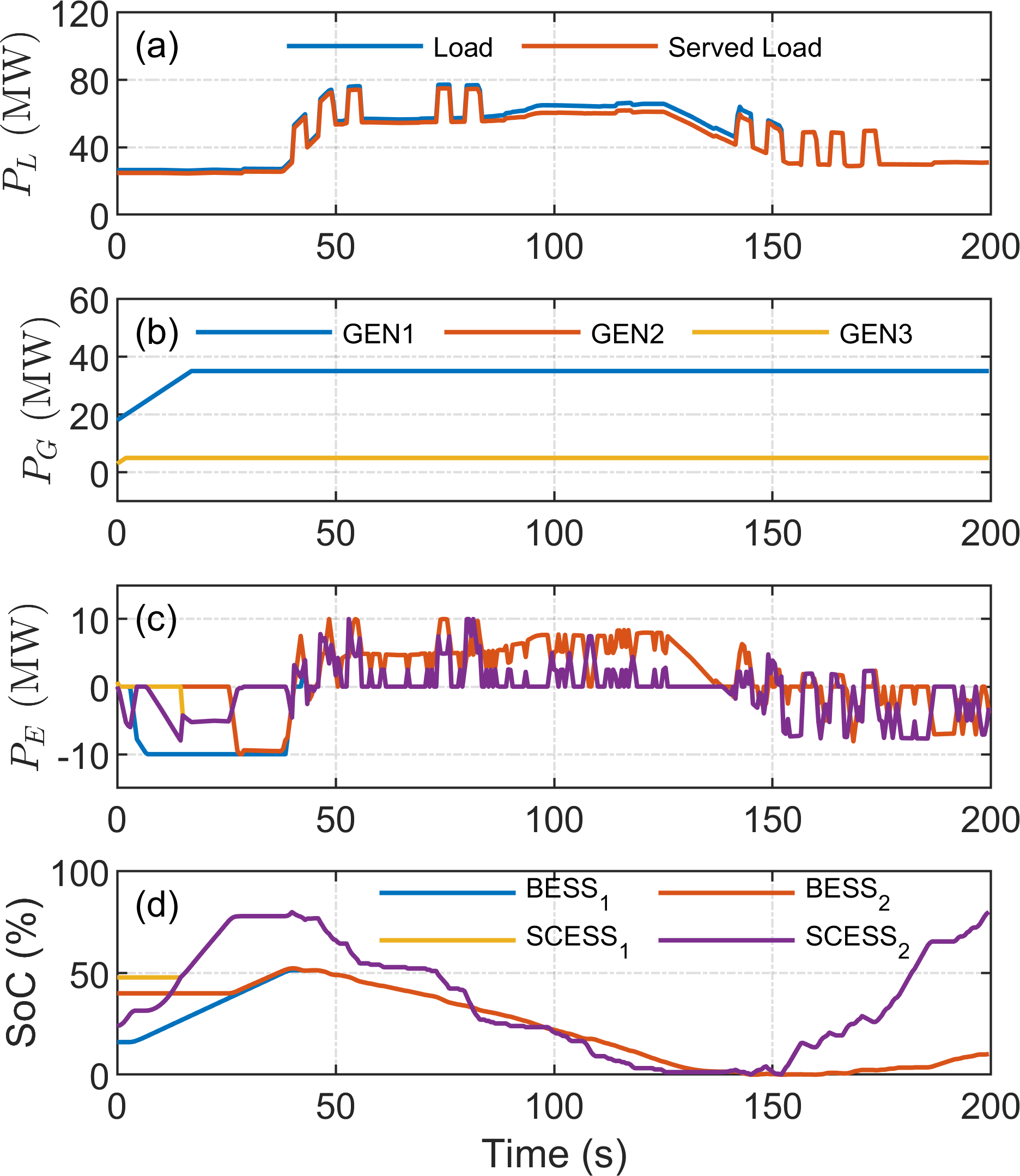}
	\caption{Optimization problem with optimal weights of secondary objectives.}
	\label{fig:LSwGD}
\end{figure}

The tested system, which consists of three generators with a total capacity of 45~MW, two units of 10MW-BESS, and two units of 10MW-SCESS, are used to evaluate the effectiveness of the proposed GD algorithm on finding optimal weights. The energy of BESS and SCESS are given in Table~\ref{TB:systemparameters}. Maximum power generation is 85~MW.  Fig.~\ref{fig:LSwoGD} shows the optimization results of the ship power system without considering secondary objective terms of $f_1, f_2,$ and $f_3$. It can be seen in Fig.~\ref{fig:LSwoGD}(a) that loads are mainly shed in the period from 80~s to 145~s. There are several issues, such as circulating power among ESS and low SoC levels of ESS at the end of optimization window, which might result in failure of ESS to serve loads in the next mission. In addition, the SoC levels are different among ESS. Fig.~\ref{fig:LSwGD} shows the optimization results considering secondary objectives with optimal constant weights. It can be seen that the above issues are addressed. Initial SoC levels of four ESS are different, however, they are equal after 30~s due to optimally scheduling ESS powers. SoC levels at the end of the optimization window are maximized to prepare for the next mission. In addition, since the SCESS are prioritized, their SoC levels are higher than those of BESS.

\section{Proposed Receding Horizon Optimization}
\label{sec:RHO}

The optimization problem from (\ref{eq:objfunc}) can be solved by the fixed horizon optimization (FHO) method, in which the optimization window is the whole mission time. Solving problem (\ref{eq:objfunc}) yields the optimal operation status of the system in all time steps. However, the resilience operation of the ship power system requires a small control time step (~0.5s) to fulfill the need of high ramp-rate loads. The required small time step causes the size of the optimization problem to increase significantly. The MILP problem creates additional challenge as they are highly nonlinear, which poses a significant impact on the computation required for real-time implementation.

To solve the above-mentioned problem, the receding horizon optimization (RHO) method is used in this paper. The RHO framework is designed with a series of shorter time windows to achieve the goal instead of one long trajectory as seen in the FHO method. The RHO-based optimization problem is solved at each time step to find a set of actions over a fixed time horizon (window), in which only first time-step solution of each window is applied to the physical system. The optimization process is repeated at the next time steps, in which a new optimization problem is solved when the time horizon shifted one step forward. The size of the MILP problem in RHO is substantially smaller than in FHO, making RHO suitable for real-time applications. The RHO method can involve real-time measurements at each time step as feedback to determine the optimal actions and even more accurate than FHO due to the system uncertainties. When the time horizon length is long enough to reach the goal, RHO is equivalent to the FHO method under the deterministic scenarios. The problem (\ref{eq:objfunc_linearize}) is represented in the RHO problem, as in (\ref{eq:objfuncRHO}).

\begin{maxi}[1]
	{x}
	{	f(x) = f_1(o_i^t) 
	- \omega_1 f_2(u_P^t) 
	}
	{}{}
	\breakObjective{- \omega_2 f_3(u_{SoC}^t) + \omega_3 f_4(SoC^{T})},
	\label{eq:objfuncRHO}
\end{maxi}
%where
\begin{align}
	x & = [o_i^t,P_e^{E,t}, P_g^{G,t}, u_P^t, u_{SoC}^t]^\top, \\
	\label{eq:objfuncf1RHO}
	f_1(o_i^t)  
	&= {
		\sum_{t \in 1}^{N_p}	
		{\sum_{i \in 1}^{n_L}{\hat{w}_i o_i^t}}},\\
	f_2(P_e^{E,t})  
	\label{eq:objfuncf2RHO}
	&= \sum_{t \in 1}^{N_p}{\sum_{e \in 1}^{n_E}{u_P^t}},\\
	f_3(SoC^t) 
	\label{eq:objfuncf3RHO}
	&= \sum_{t \in 1}^{N_p}{\sum_{l,m}{u_{SoC}^t}},\\
	f_4(SoC^{T}) 
	\label{eq:objfuncf4RHO}
	&= \sum_{e \in 1}^{n_E
	}{\alpha_e SoC_e^{E,T}},	
\end{align}
where $N_p$ is the horizon length. 

\section{Case Studies}
\label{sec: casestudies}

\begin{figure*}[t]
	\centering
	\includegraphics[width=0.9\linewidth]{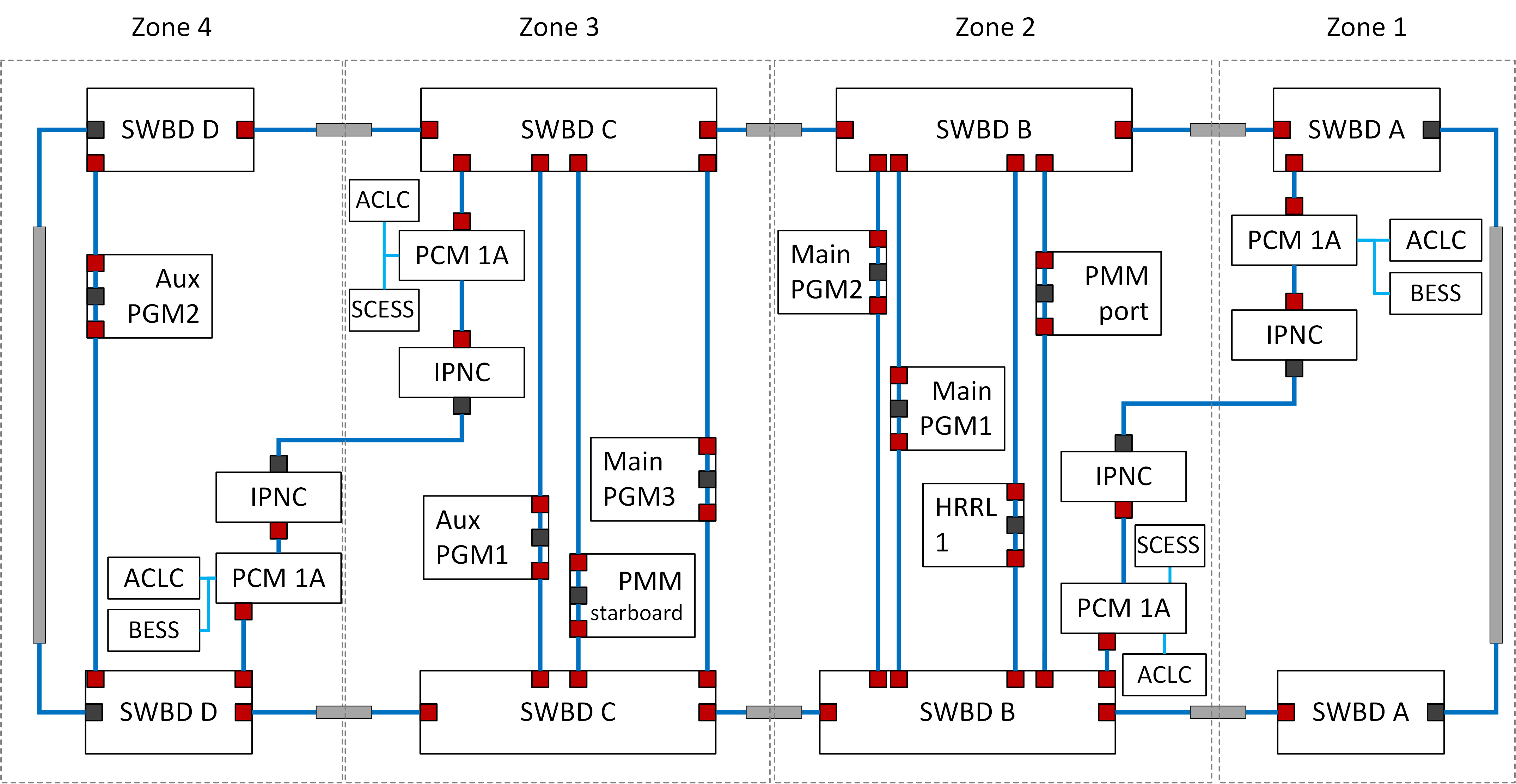}
	\caption{Notional four-zone MVDC shipboard power system adapted from \cite{MVDC2020model}.}
	\label{fig:MVDC_Notional}
\end{figure*}

RHO is used to manage the ship power system under the condition of high ramp-rate load. A comparison with the FHO is presented in this section to show the effectiveness of the proposed RHO method. 

\subsection{System Description}

The notional four-zone MVDC ship power system shown in Fig.~\ref{fig:MVDC_Notional} is used to evaluate the performance of the proposed EMS. The power rating of the shipboard power system is 100~MW. Each zone is composed of multiple modules such as power conversion module (PCM-1A), power generation module (PGM), propulsion motor module (PMM), integrated power node center (IPNC), and energy storage module (ESM). Parameters of the energy storage systems are given in Table~\ref{TB:systemparameters} and the RHO parameters are given in Table~\ref{TB:controlparameters}. More detailed information on the MVDC shipboard power system can be found in \cite{MVDC2020model}.

%https://www.esrdc.com/media/1088/fzmvdc_mdd_v30_43-7281-20.pdf

\begin{table}[h]
	\caption{RHO parameters}
	\label{TB:controlparameters}
	\begin{tabularx}{3.0in}{c|c|c}
		\toprule
		Symbol &  Parameter & Value\\ % Table header row
		\midrule
		$\Delta t$ 		&	Control sample time		& 0.5~s \\
		$N_p$ 			& 	Horizon length				& 60 \\
		$T$ 			& 	Mission time				& 600~s \\
		\bottomrule
	\end{tabularx}
\end{table}

\subsection{Performance Evaluation}

\subsubsection{Performance under High ramp-rate Load Mission}
\begin{figure}[t]
	\centering
	\includegraphics[width=0.9\linewidth]{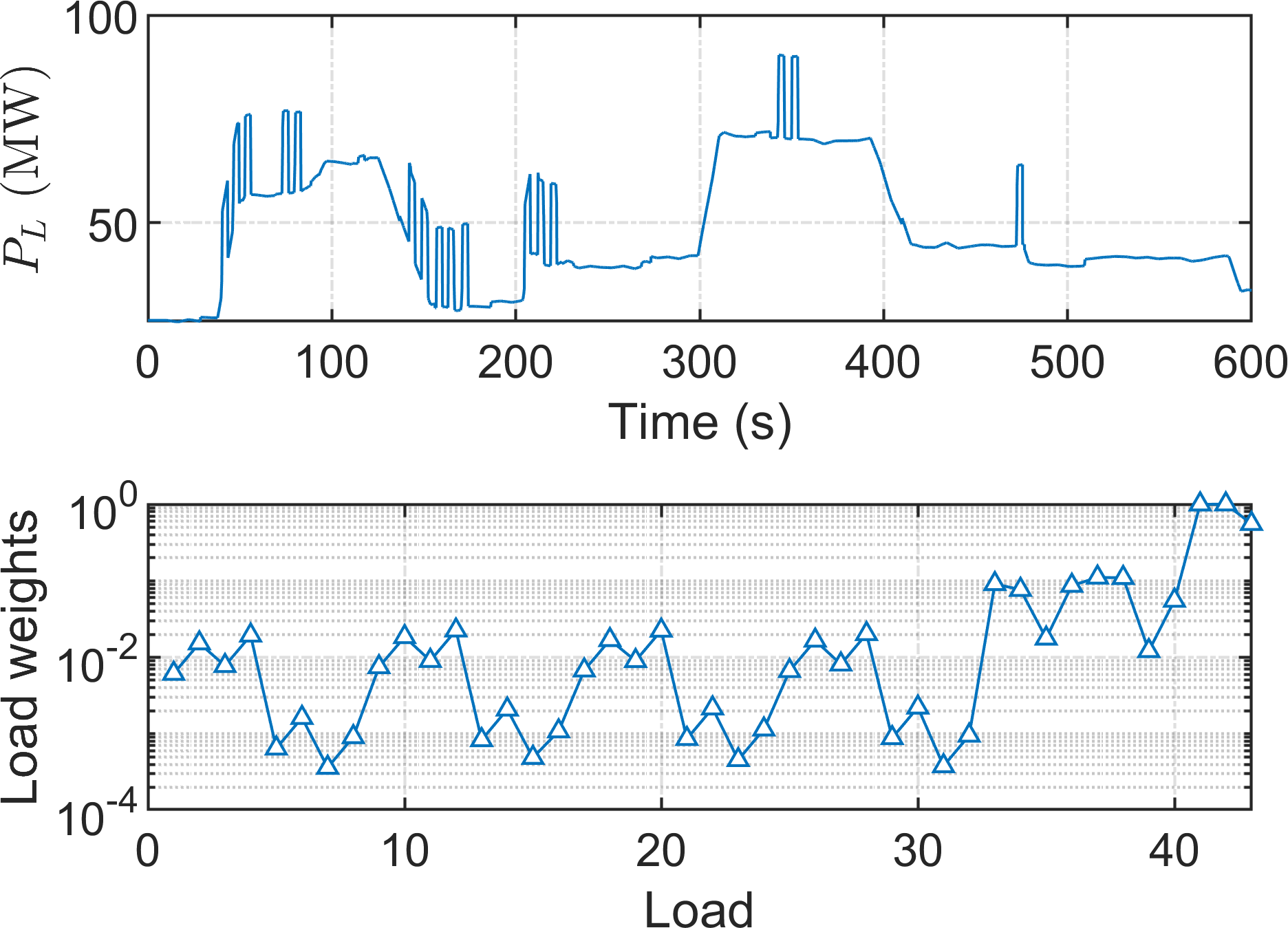}
	\caption{Load profile and weights of 43 loads.}
	\label{fig:Loaddata}
\end{figure}

The proposed RHO is evaluated in the condition of HRRL operation. In this condition, the ramp-rate of load power may exceed generation ramping. The uses of ESS, particularly SCESS, can support the ship power system in this condition. The system 43 loads' profile and their weights are shown in Fig.~\ref{fig:Loaddata}. 

\begin{figure}[t]
	\centering
	\subfigure[All loads]
	{
		\includegraphics[width=0.9\linewidth]{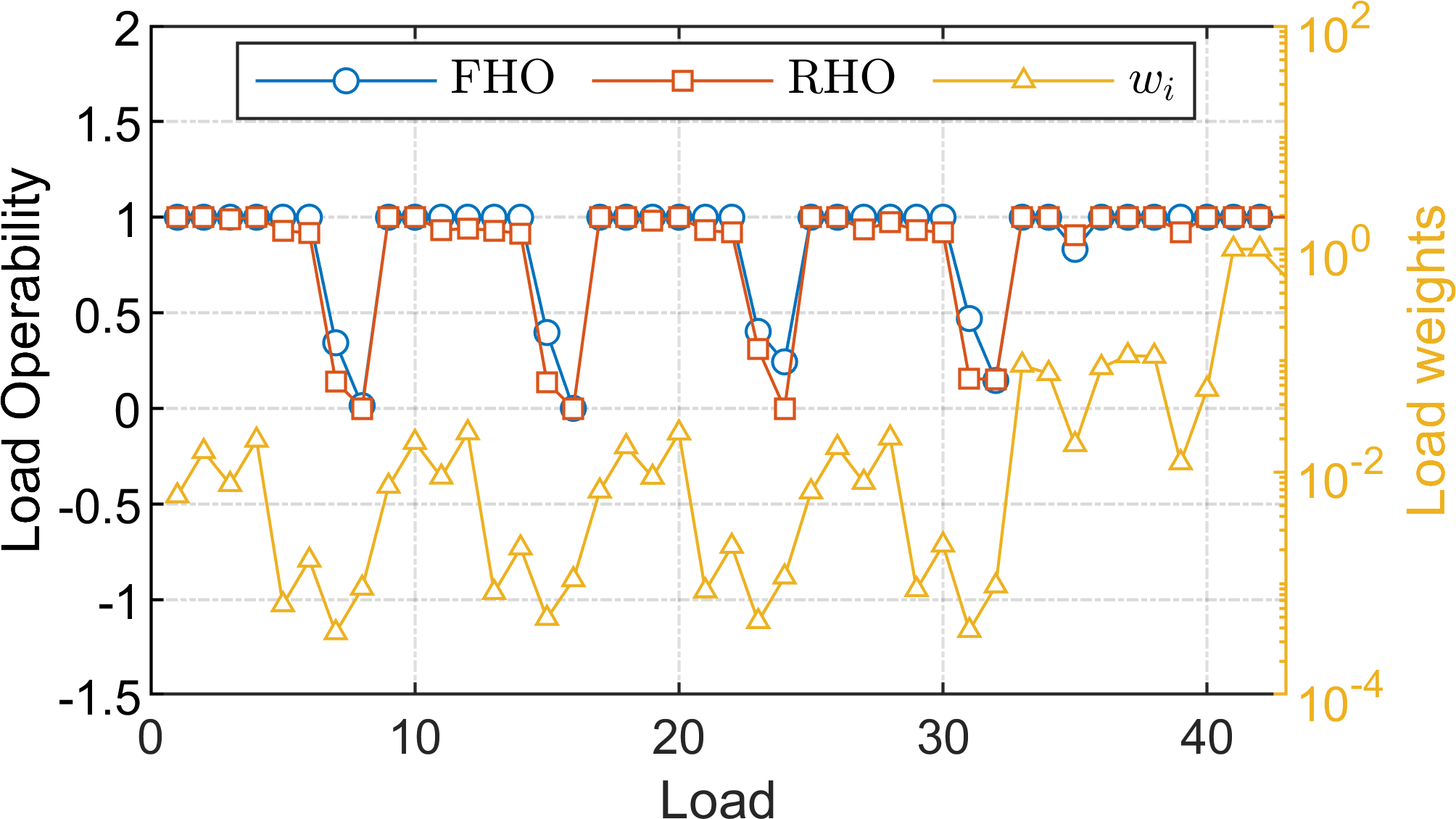}
	} \quad	
	\subfigure[Loads 4 to 10]
	{
		\includegraphics[width=0.9\linewidth]{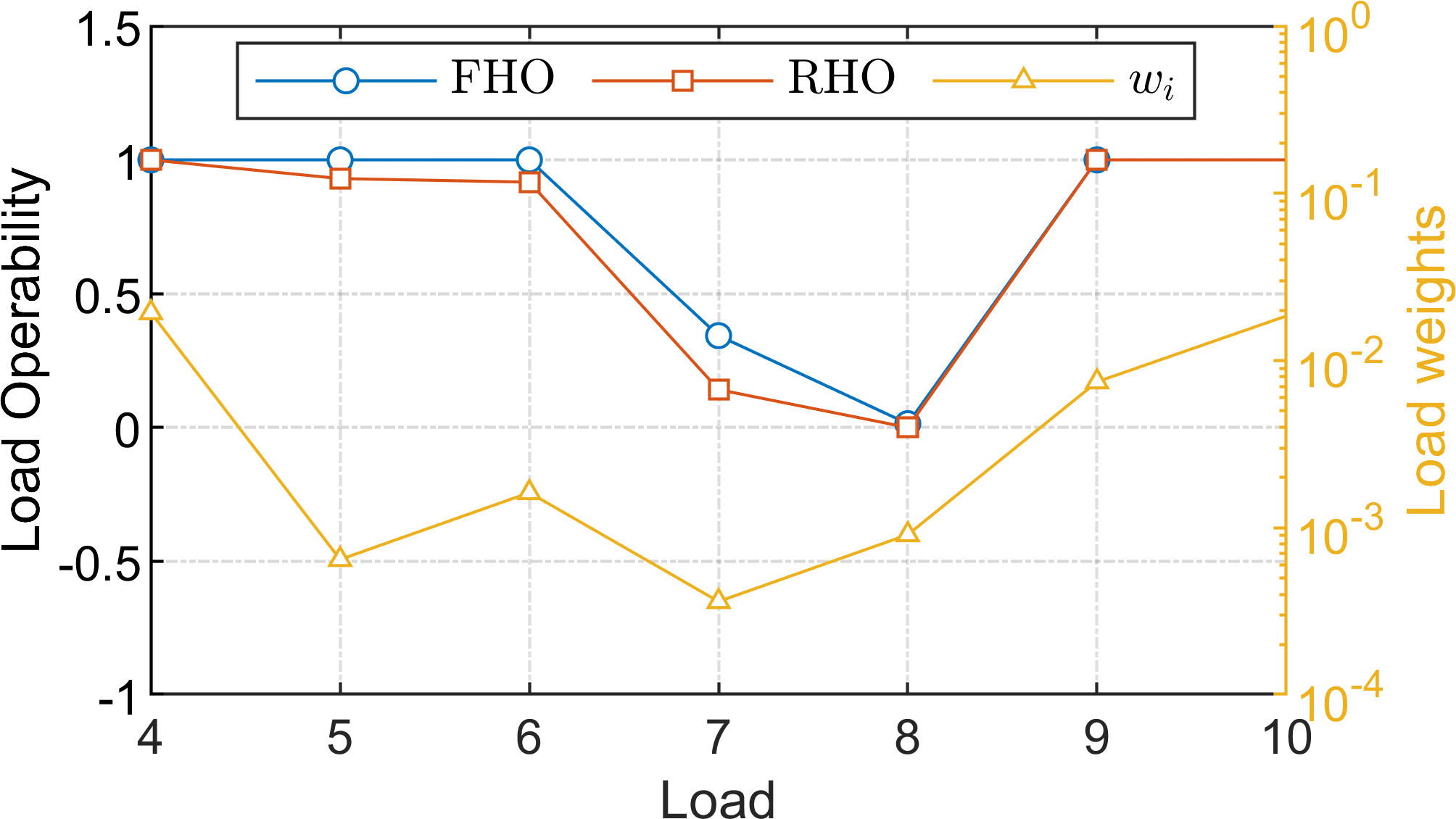}
	} 
	\caption{Compared load operability in HRRL mission.}
	\label{fig:Operability}
\end{figure}

\begin{figure}[t]
	\centering
	\subfigure[Whole mission]
	{
		\includegraphics[width=0.9\linewidth]{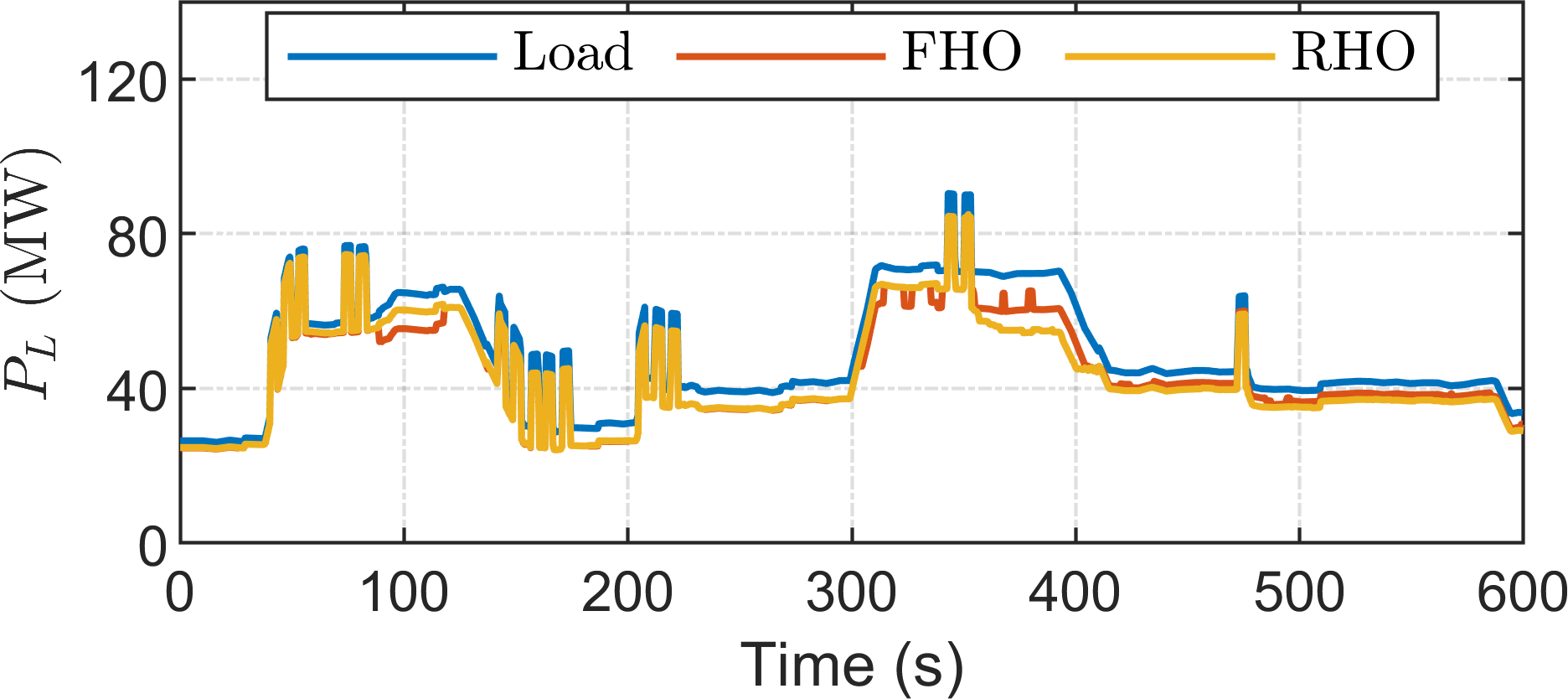}
	} \quad	
	\subfigure[Zoon in]
	{
		\includegraphics[width=0.9\linewidth]{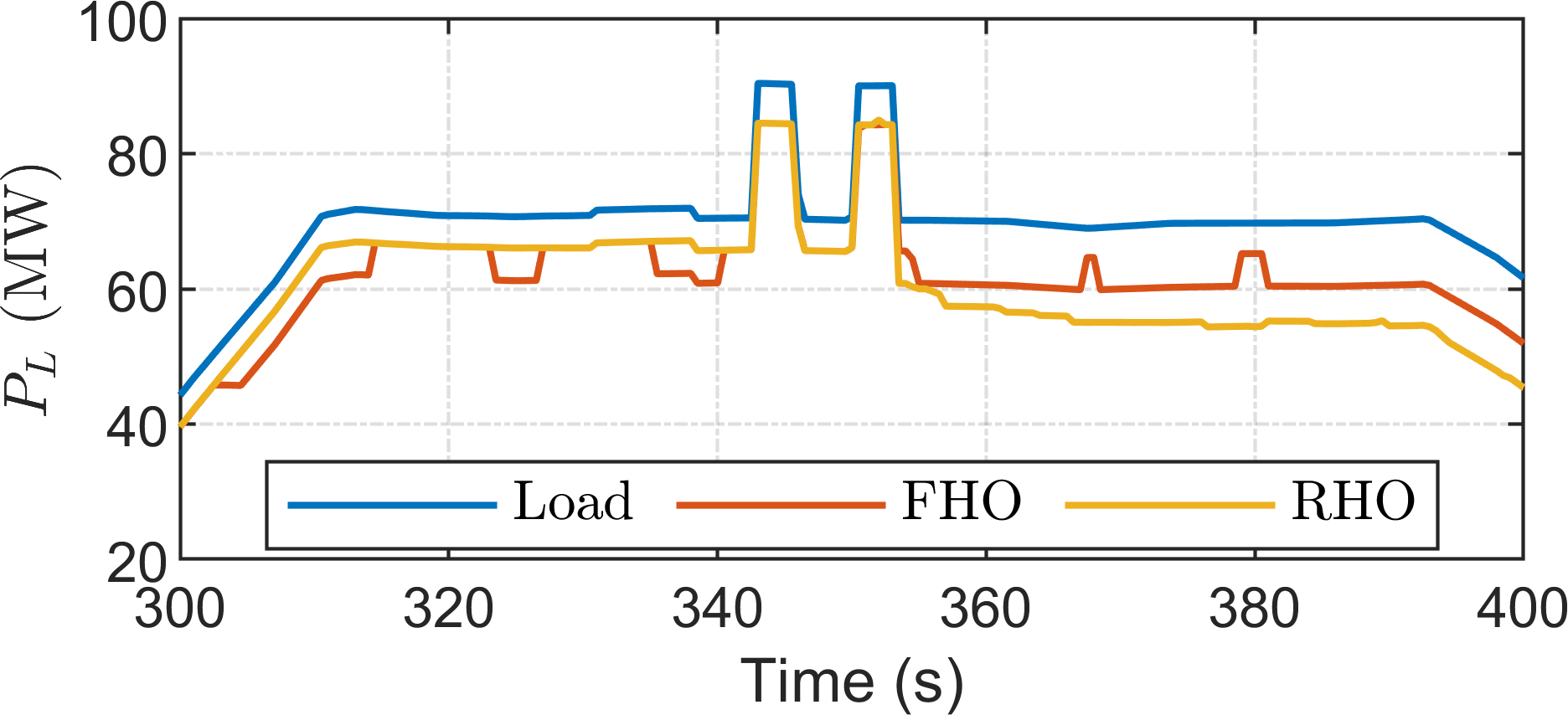}
	} 
	\caption{Load profile and served load under RHO and FHO methods.}
	\label{fig:Servedload}
\end{figure}

The HRRL loads (load numbers 41, 42, and 43) have the largest weight values, which must be given top priority.  Fig.~\ref{fig:Operability}(a) shows the load weights and optimal load operability that is found by two methods (RHO and FHO). When the load operability value is less than one, the load is shed. It can be seen that loads with low weight $\hat{w}_i$ are shed while HRRL with the highest weight is served, which is depicted in Fig.~\ref{fig:Operability}(b). Fig.~\ref{fig:Servedload}(a) shows the profiles of served load under two optimization methods. With the FHO method, more loads are shed before 300~s to serve for the increase in load demand from 300~s to 400~s, as depicted in Fig.~\ref{fig:Servedload}(b). In the case of RHO, more loads are shed in period from 350~s to 400~s compared to FHO as RHO has shorter optimization length compared to FHO. The error of total load operability between two methods given by (\ref{eq:deltaf}) indicates that the performance of the proposed RHO is close to FHO. 

\begin{align}
	\label{eq:deltaf}
	\Delta f_1 = {{f_1^{FHO} - f_1^{RHO}}\over {f_1^{FHO}}} = +0.05\%
\end{align}
\begin{figure}[t]
	\centering
	\includegraphics[width=0.9\linewidth]{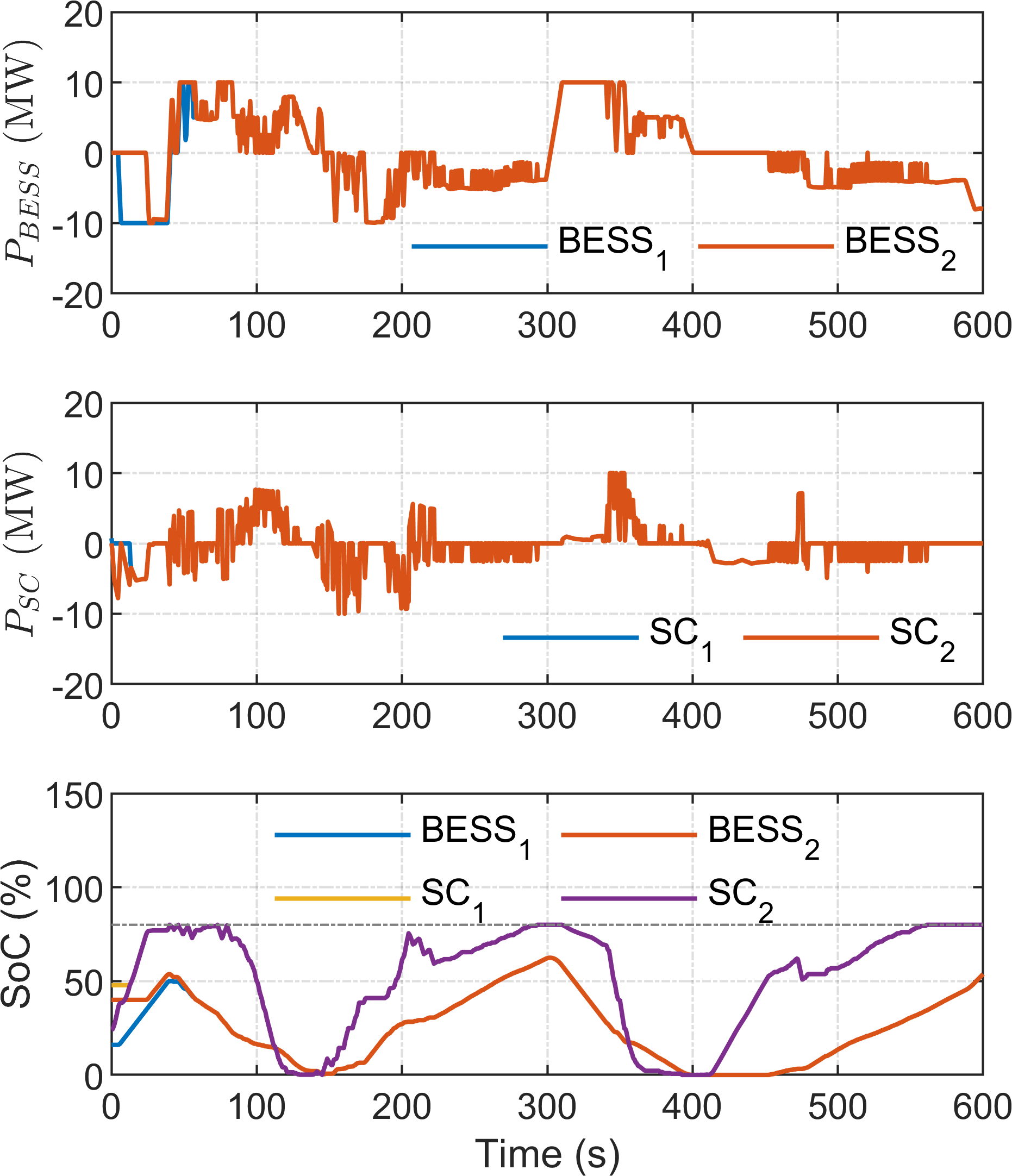}
	\caption{ESS power and energy when using RHO.}
	\label{fig:RHO_ESS}
\end{figure}

\begin{figure}[t]
	\centering
	\includegraphics[width=0.9\linewidth]{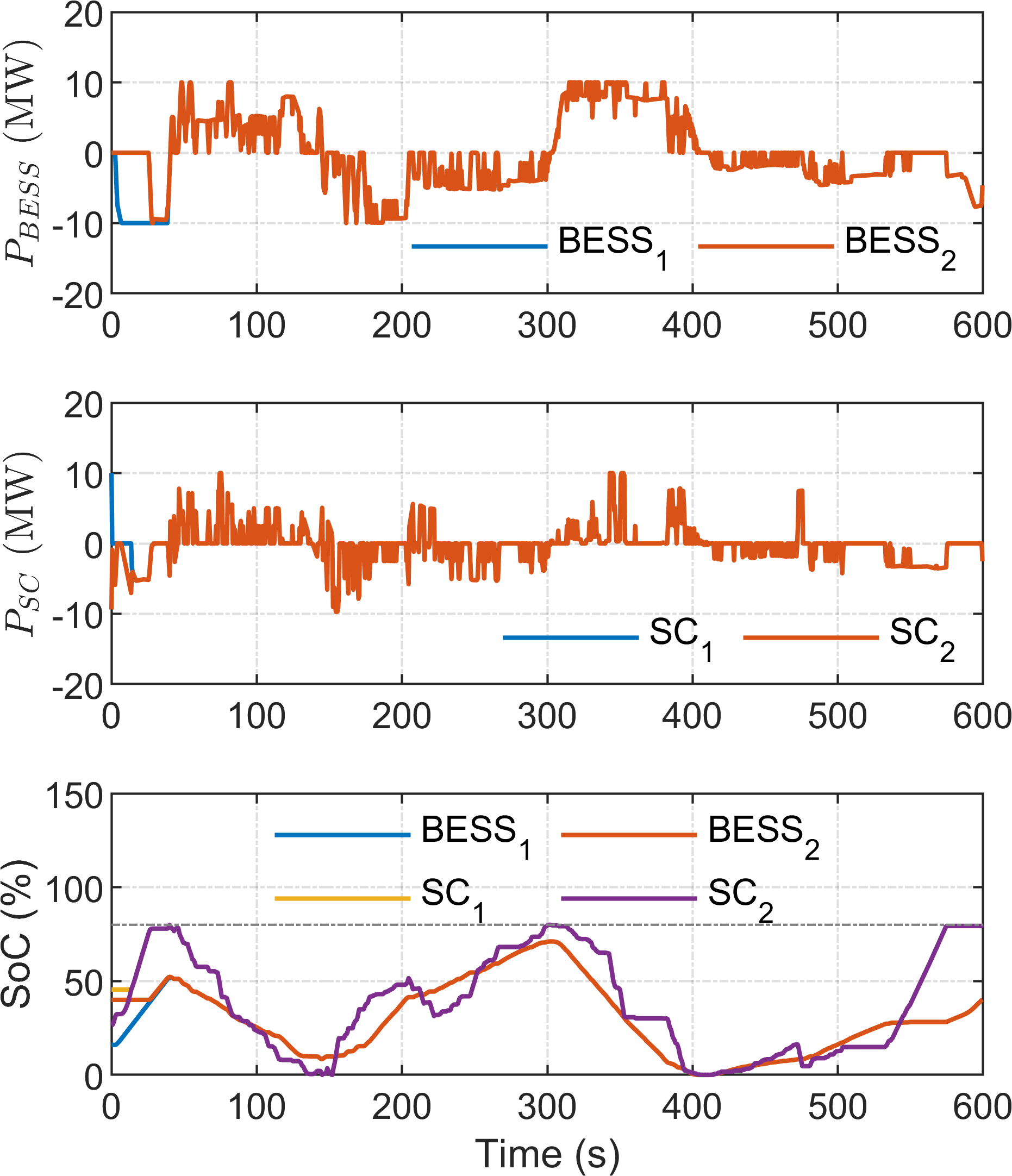}
	\caption{ESS power and energy when using FHO}
	\label{fig:FHO_ESS}
\end{figure}

The output power of BESS and SCESS are shown in Figs.~\ref{fig:RHO_ESS} and~\ref{fig:FHO_ESS}. From 0~s to 50~s, four ESS are charged as the generation is larger than load demand. Since ESS have different SoC levels, the charging power of each ESS is different accordingly to balance SoC. It can be seen that two BESS and two SCESS have the same SoC levels after 50~s and the output power of either two BESS or two SCESS are the same. The maximum SoC level of ESS is kept at 80\%. For both methods, SCESS are given priority to charge, resulting in higher SoC levels at the end of the optimization window compared to BESS. It is observed that the proposed RHO method achieves a similar performance yielded by the FHO method.

%\begin{figure}[h]
%	\centering
%	\includegraphics[width=0.9\linewidth]{SPO_ESSzoom.png}
%	\caption{Compared load shedding in HRRL mission (Zoom in)}
%	\label{fig:SPO_ESSzoom}
%\end{figure}

\subsubsection{RHO Performance in Different Scenarios}

\begin{figure}[t]
	\centering
	\includegraphics[width=0.9\linewidth]{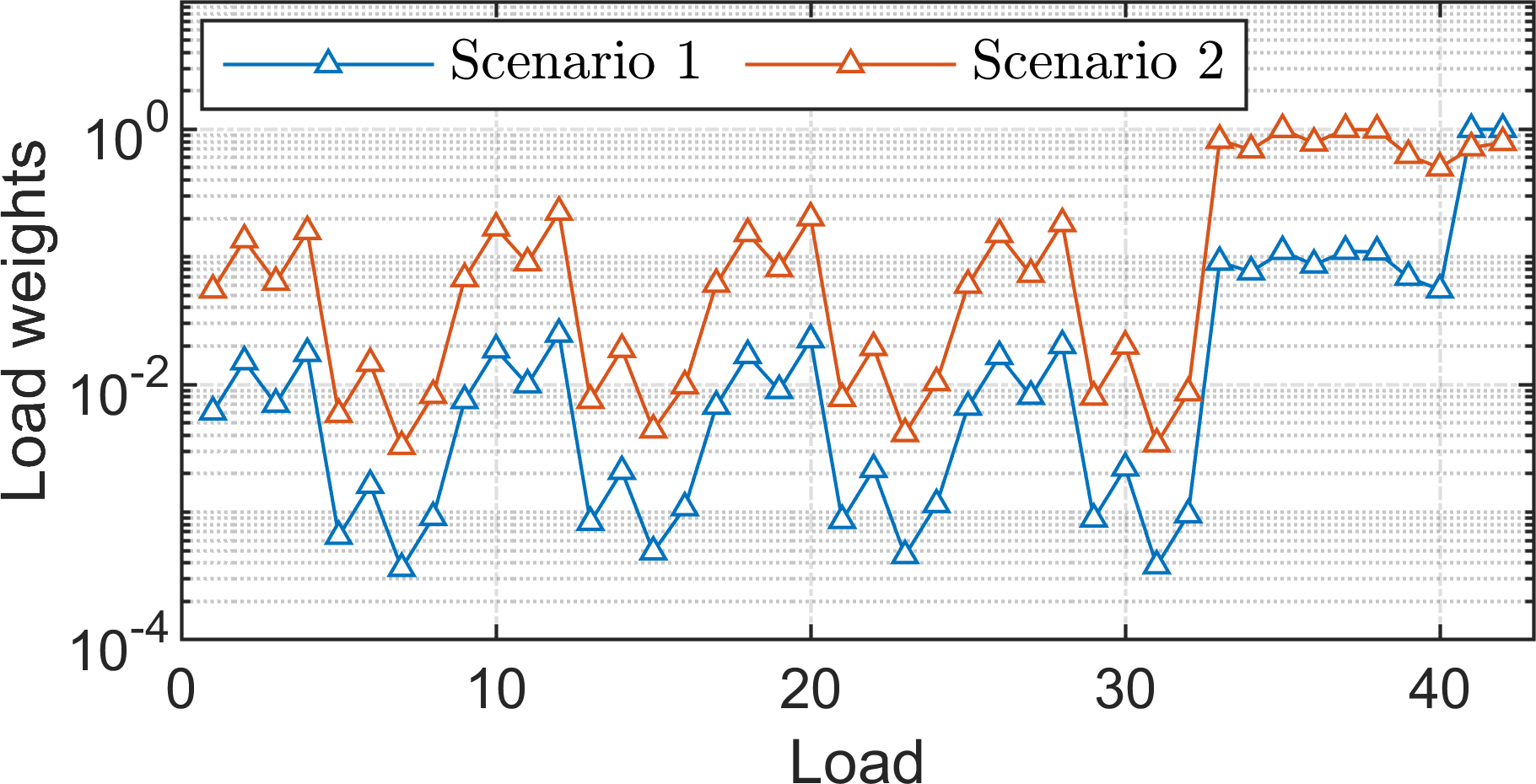}
	\caption{Weights of loads for different scenarios.}
	\label{fig:Lweights}
\end{figure}

The MVDC ship power system is operated under various scenarios such as peacetime cruise, sprint station, battle, and anchor \cite{MVDC2017model}. The importance of loads varies with the operating conditions of the ship power system. A load might be vital in one scenario but not in another. For example, when the mission of the ship power system changes from cruise to battle, some loads that were previously non-vital become vital. The proposed RHO method is evaluated by two additional scenarios, as shown in Fig. \ref{fig:Lweights}.
\begin{figure}[t]
	\centering
	\includegraphics[width=0.9\linewidth]{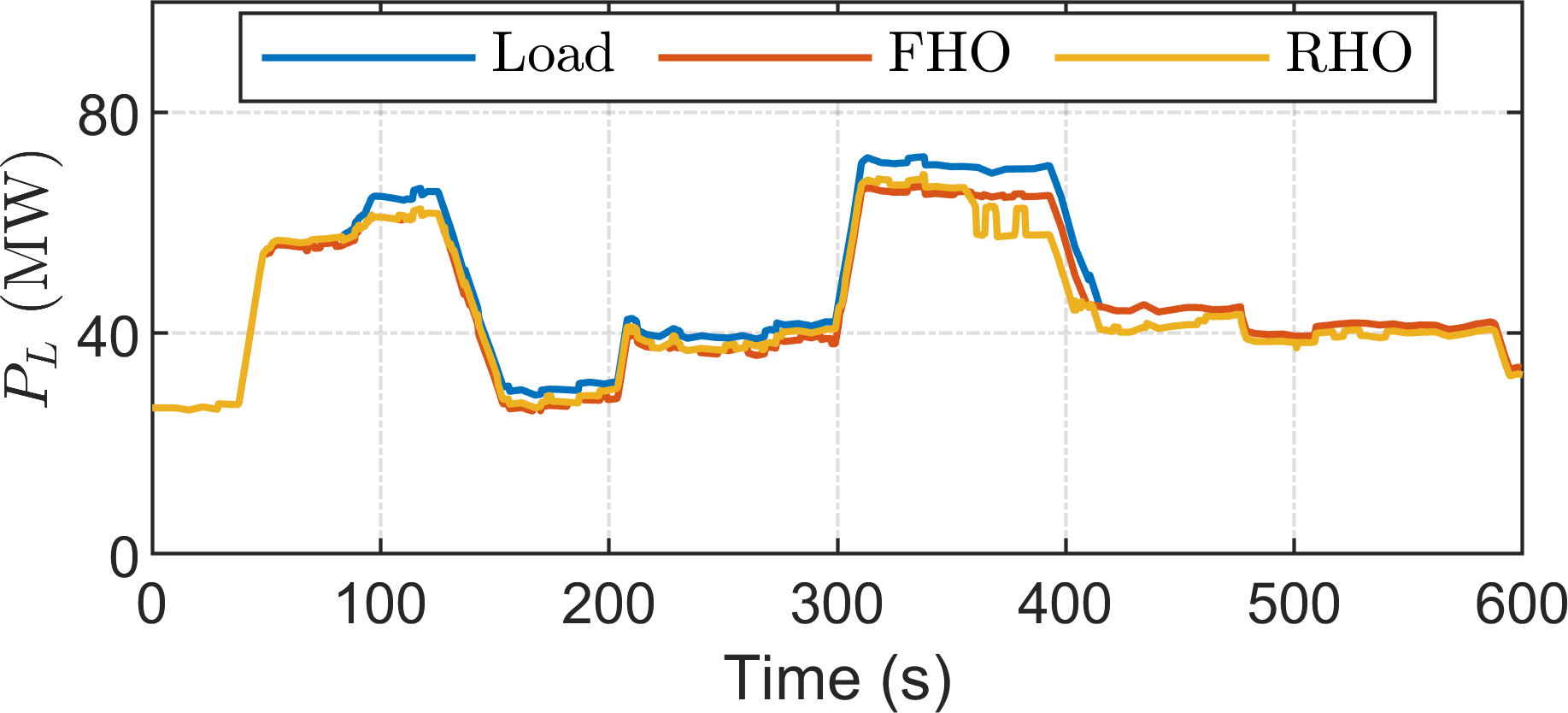}
	\caption{Served load in scenario 1.}
	\label{fig:LSbased}
\end{figure}
\begin{figure}[t]
	\centering
	\includegraphics[width=0.9\linewidth]{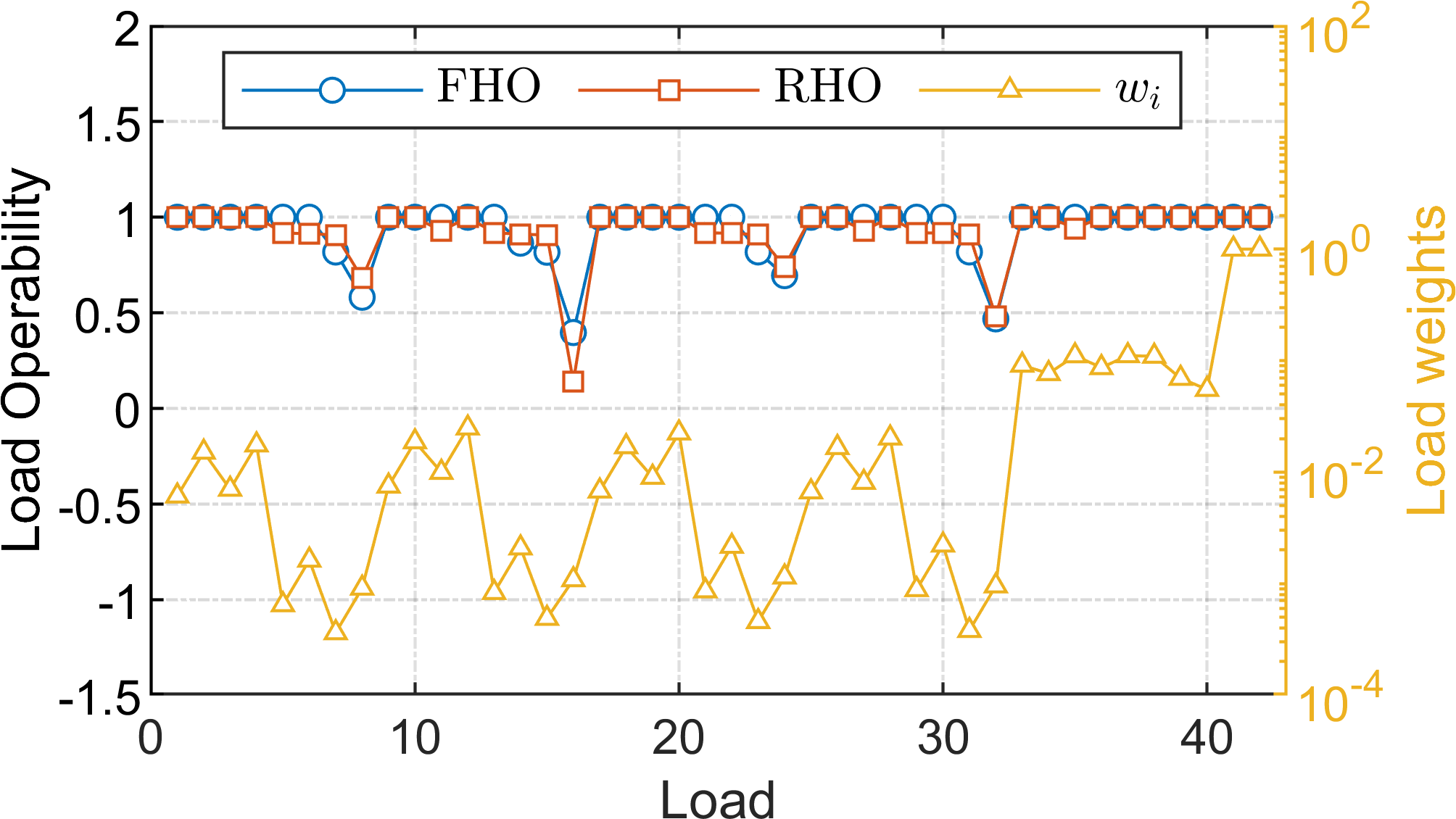}
	\caption{Compare load operability in scenario 1.}
	\label{fig:LSbasedOi}
\end{figure}
\begin{figure}[h]
	\centering
	\includegraphics[width=0.9\linewidth]{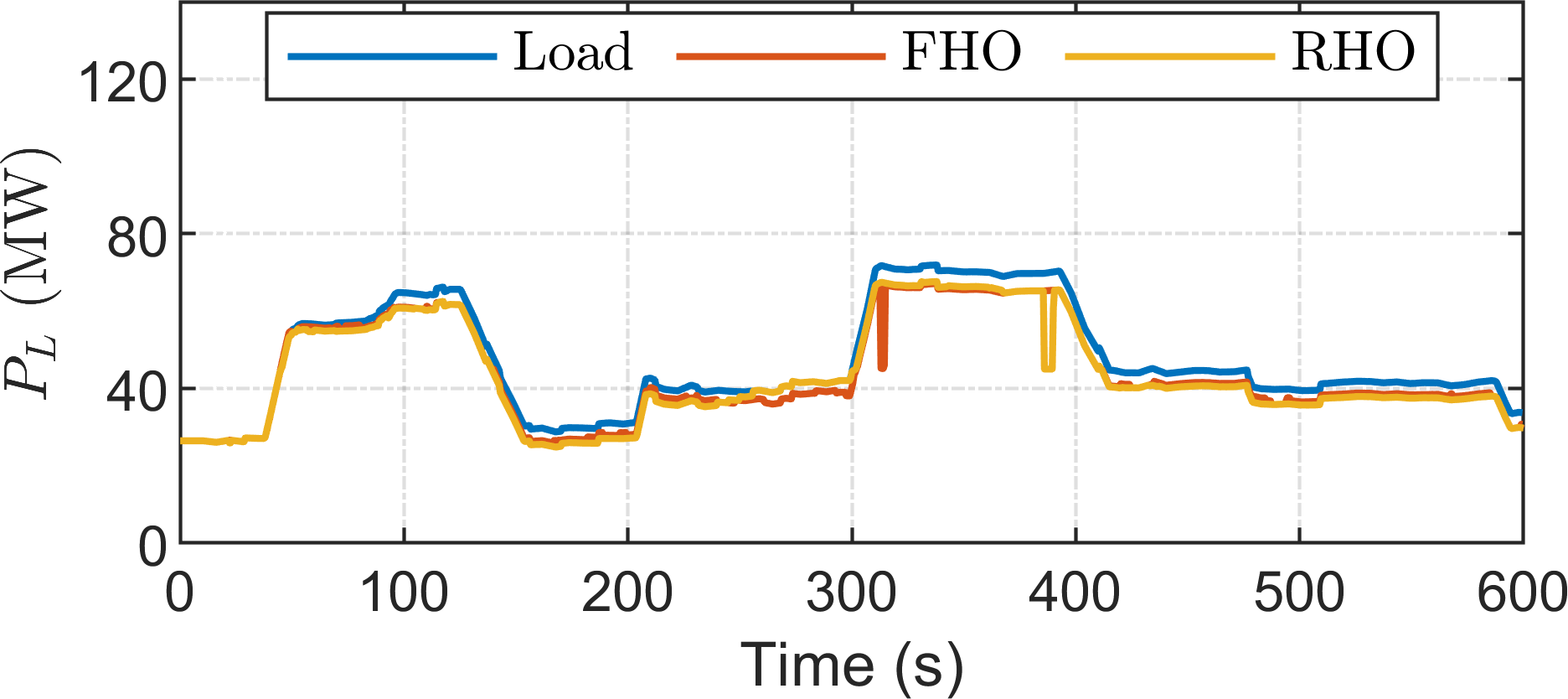}
	\caption{Served Load in scenario 2.}
	\label{fig:LSlowPMM}
\end{figure}
\begin{figure}[h]
	\centering
	\includegraphics[width=0.9\linewidth]{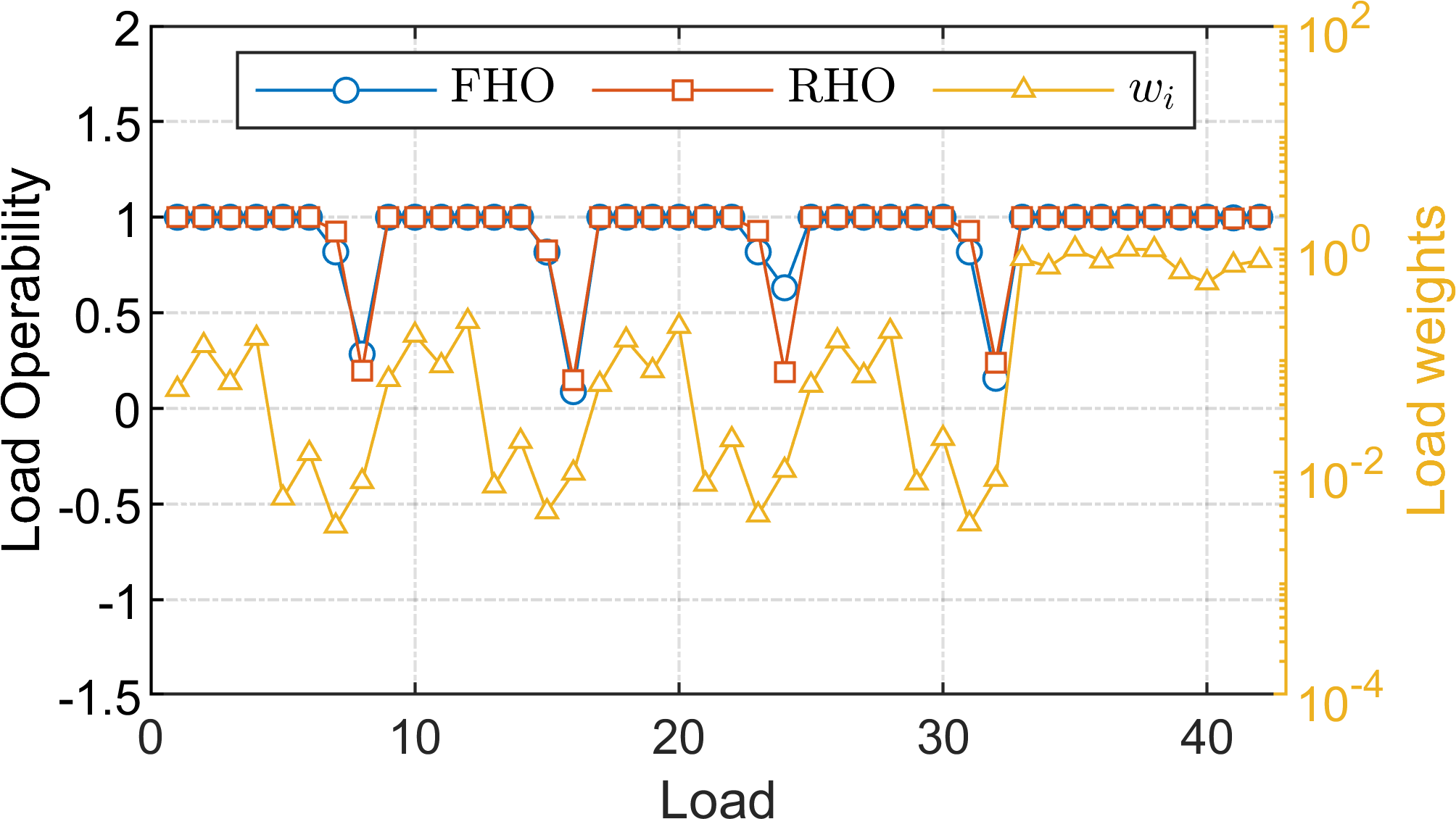}
	\caption{Compare load operability in scenario 2.}
	\label{fig:LSlowPMMOi}
\end{figure}

Figs.~\ref{fig:LSbased} and \ref{fig:LSbasedOi} show the comparison between proposed RHO and FHO methods for the first scenario. Loads are mainly shed from 50~s to 400~s and only loads with low weight values are mainly shed, as shown in Fig.~\ref{fig:LSbasedOi}. Figs.~\ref{fig:LSlowPMM} and \ref{fig:LSlowPMMOi} show the performance of RHO method compared to FHO method for the second scenario. In this case, in stead of shedding significant load at 314~s (FHO), RHO method sheds significant loads at 385~s. As the optimization window of RHO method is much shorter than that of FHO method, loads  are still served at 314~s. However, all ESS are almost fully discharged after serving loads from 314~s to 384~s, loads are then shed significantly at 385~s due to the shortage of power generations. The total load operability shown in Figs. \ref{fig:LSbasedOi} and \ref{fig:LSlowPMMOi} indicates which loads are mainly shed in two scenarios. Non-vital loads with low weight values are primarily shed in both cases, whereas the vital loads with high weight values are served. Overall, it can be observed that the proposed RHO method achieves a similar performance compared to the FHO method.

\begin{comment}

\subsubsection{Effect of Horizon Length on the Optimization Performance}

{\color{red} Add simulation results.}

\end{comment}

\subsection{Computational Comparison}

\begin{figure}[t]
	\centering
	\includegraphics[width=0.9\linewidth]{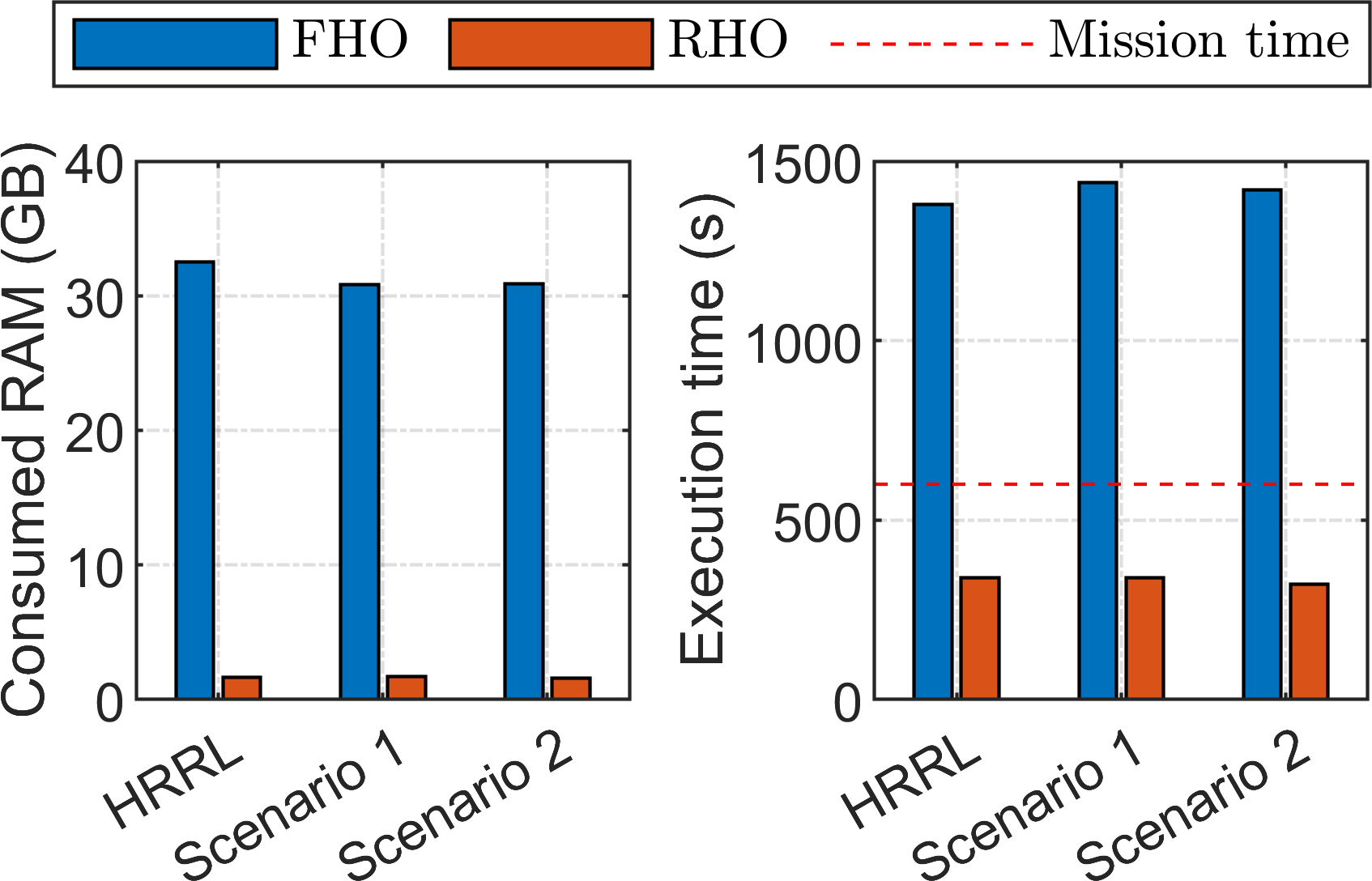}
	\caption{Computational comparison between RHO and FHO. Execution time in case of RHO method is the sum of execution time for all steps in 600s-mission.}
	\label{fig:LScomputation}
\end{figure}

\begin{figure}[t]
	\centering
	\includegraphics[width=0.9\linewidth]{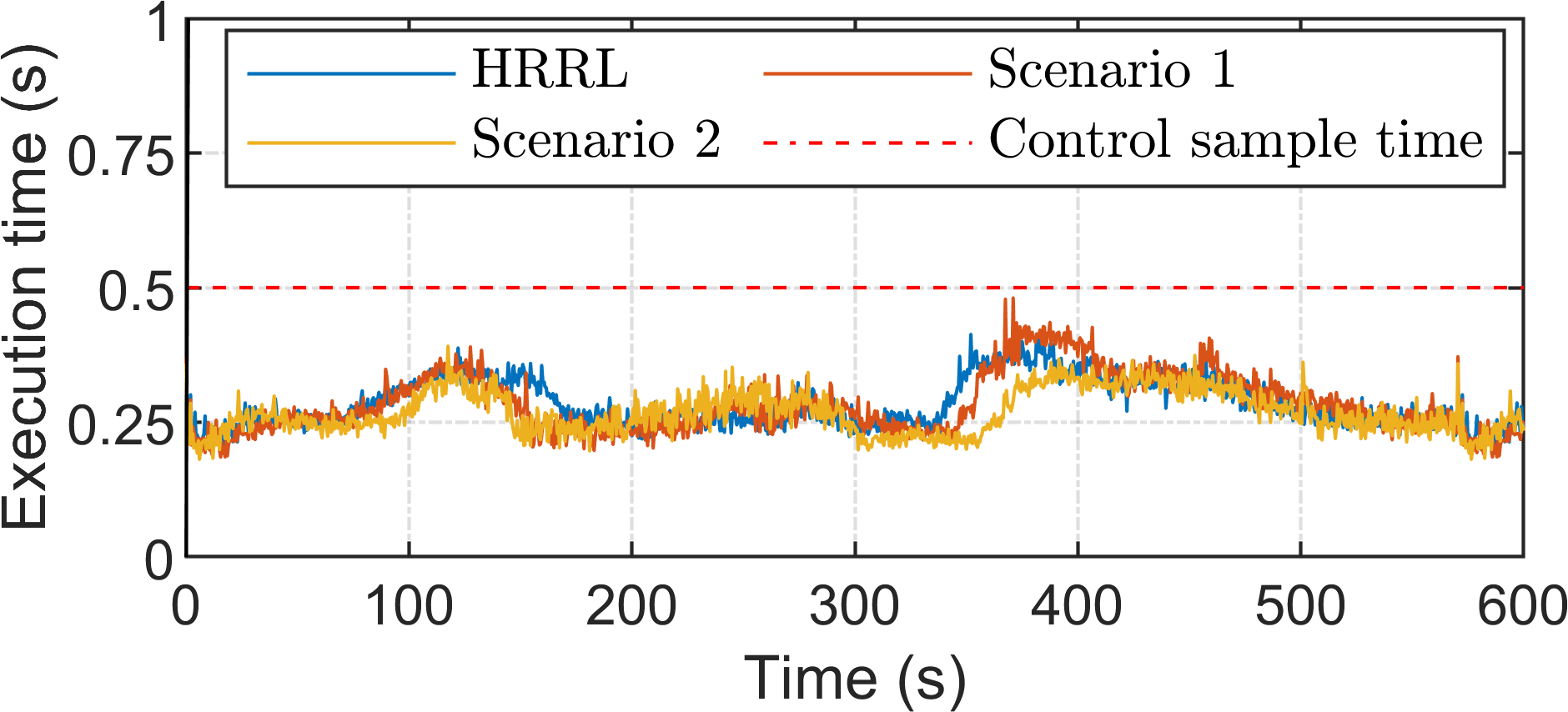}
	\caption{Execution time of the RHO method measured at each optimization step.}
	\label{fig:LSexecutetime}
\end{figure}

The above sections presented two methods that achieve similar performance under various scenarios. A comparison between proposed RHO and FHO in Fig.~\ref{fig:LScomputation} show the advantage of proposed RHO method in terms of computation. Both methods are performed on a processor of Intel$\circledR$Core~i7-10700H at 2.90~GHz and 64~GB RAM. The \textit{intlinprog} function based on \textit{primal-simplex} algorithm from MATLAB is used to solve the MILP problem. Various scenarios are conducted to evaluate the computational performance of both methods, such as HRRL, scenarios 1, and 2. In all scenarios, the total mission time is~600~s. In FHO method, the required RAM to solve the optimization problem is about 31$\sim$33~GB, wheres 1.6$\sim$1.7~GB RAM are required for RHO method. Despite using a powerful computer, FHO takes 1380$\sim$1440~s to solve the optimization problem. To ensure that the optimal solution can be found before applying the control actions, the execution time must be less than 600~s in the case of FHO, and in the case of RHO, the execution time at each step must be less than the control sample time of 0.5~s. It can be seen that the FHO method is not applicable for the real-time implementation of the required 600~s. In comparison, the proposed RHO method completes the optimization problem in less than 0.5~s, as shown in Fig. \ref{fig:LSexecutetime}. Therefore, the RHO method is applicable for the real-time implementation. Compared to the FHO method, the proposed RHO method shows significant improvements in terms of computation as its RAM usage is 95\% lower and its execution time is 76\% faster. 

%The proposed RHO method shows significant improvements in terms of computation as it requires much lower RAM, and its execution time is much shorter compared to the FHO method {\color{red} quantify the percentages/ number of times in improvements}.

%\begin{table}[h]
%	\caption{Computational Comparison}
%	\label{TB:computation}
%	\begin{tabularx}{3.3in}{c|c|c|c}
%		\toprule
%		Method &  Consumed RAM & Solver's time & Mission time\\ % Table header row
%		\midrule
%		FHO 		&	34~GB		& 1676~s & 600~s \\
%		RHO HRRL		&	1.67~GB	& 340~s & 600~s\\
%		RHO 	s1	&	1.67~GB	& 339~s & 600~s\\
%		RHO 	s2	&	1.67~GB	& 322~s & 600~s\\
%		
%		\bottomrule
%	\end{tabularx}
%\end{table}

\section{Conclusion}
\label{sec:conclusion}

A resilience-oriented EMS of ship power systems has been proposed in this paper, which manages multiple types of energy storage systems such as BESS and SCESS to maximize load operability. In addition to enhancing system resilience, secondary objectives of managing ESS were involved, such as balancing SoC among ESS and prioritizing the SoC level of SCESS. Since the power ramp-rate of SCESS is much higher than that of BESS, SCESS were given priority to serve HRRL. The gradient descent algorithm was proposed to optimally design the weights of secondary objectives. The RHO technique was proposed to reduce the computational burden, making it suitable for real-time applications. A comparison between RHO and FHO methods revealed that they produced comparable resilient results. However, the use of the RHO method showed a significant improvement in computation, with significantly less RAM consumption and execution time. The paper assumed that the power cables have enough capacity to transfer power among subsystems. Our future research will consider the power line limits and deploy the proposed EMS in real-time controllers.

%\section*{Acknowledgment}

%{This material is based upon research supported by, or in part by, the U.S. Office of Naval Research under award number N00014-16-1-2956.}


\bibliographystyle{cas-model2-names}
\bibliography{references}

\end{document}